\documentclass[12pt]{article}
\usepackage{amsmath}
\usepackage{graphicx}
\usepackage{enumerate}
\usepackage{natbib}
\usepackage{url} 
\usepackage{bm}
\usepackage{amssymb}
\usepackage{color}
\usepackage{enumitem}
\usepackage[linesnumbered,ruled]{algorithm2e}
\newcommand{\blind}{1}

\begin{document}

\def\spacingset#1{\renewcommand{\baselinestretch}%
{#1}\small\normalsize} \spacingset{1}


\if1\blind
{
  \title{\bf On Reasonable Uses of the Restricted Spatial Regression Model: Selection Criteria, Robust Adjustments, and Computational Advances}
  \author{Jonathan R Bradley\thanks{
    The author gratefully acknowledge \textit{NSF-DMS 2547531}}\hspace{.2cm}\\
   Department of Statistics and Data Science,  University of Missouri,\\ Columbia, MO, USA\\
   Email: bradleyjr@missouri.edu}
   \date{}
  \maketitle
} \fi

\if0\blind
{
  \bigskip
  \bigskip
  \bigskip
  \begin{center}
    {\LARGE\bf On Reasonable Uses of the Restricted Spatial Regression Model: Selection Criteria, Robust Adjustments, and Computational Advances}
\end{center}
  \medskip
} \fi

\bigskip
\begin{abstract}
{The spatial linear mixed model (SLMM) consists of fixed and spatial random effects that may be effected by unmeasured confounding and collinearity. Partially motivated as a means to address potential issues with confounding, the Restricted spatial regression (RSR) model restricts spatial random effects to be in the orthogonal column space of the covariates. Recent articles have shown that the misspecified Bayesian RSR generally performs worse than the SLMM when the data is generated from the SLMM. However, we show that the misspecified Bayesian RSR model's marginal posterior distribution is equivalent up to a reparameterization to that of the SLMM's marginal posterior distribution, under a certain prior assumption on the orthogonalized regression coefficients. These equivalence relationships show that when comparing the SLMM and RSR using the Bayes factor, stictly proper scoring rules, cross-validation, and many other standard impartial selection criteria we see that the Bayesian SLMM and the misspecified Bayesian RSR are seen as equivalent under certain prior specifications. We provide several new directions for the RSR including results that show that an augmented RSR can produce clear computational advantages via a direct sampler from the posterior distribution for all hyperparameters, fixed effects, and random effects leading to implementation strategies for the SLMM via these equivalence relationships. Additionally, a type of transfer learning approach offers a new interpretation to orthogonalized regression coefficients, which we show empirically can improve inference on linearly dependent regression coefficients, via a bias adjustment, in the presence of spatial confounding. Simulations and an illustration using COVID-19 mortality data are provided.}
\end{abstract}

\noindent%
\vfill

\newpage
\spacingset{1.9} 
\section{Introduction}
\label{sec:intro}

The spatial linear mixed model (SLMM) is covered in nearly every standard modern spatial statistics textbooks \citep[][among several others]{diggle,cressie-wikle-book,banerjee-etal-2004}. We start with a formal statement of the SLMM as follows,
\begin{align}\label{eq:slmm}
	\textbf{y} &= \textbf{X}\bm{\beta} + \bm{\Phi}\bm{\nu}+ \bm{\epsilon},
\end{align}
\noindent
where $\textbf{y}$ is an $n$-dimensional real-valued data vector, whose $i$-th element often represents a spatially referenced response at the $i$-th observed location in the spatial domain, $\textbf{X}$ is a $n\times p$ matrix of known covariates, the ``linearly dependent regression effect'' $\bm{\beta}\in \mathbb{R}^{p}$ is unknown, $\bm{\Phi}$ is a known $n\times r$ matrix of basis functions, and $\bm{\nu}$ is a mean-zero $r$-dimensional normally distributed random vector with covariance matrix $\bm{\Sigma}= \mathrm{cov}(\bm{\nu})$. We allow for the case where $r = n$ and $\bm{\Phi} = \textbf{I}$, where $\textbf{I}$ is an $n\times n$ identity matrix. The vector $\bm{\beta}$ is referred to as ``linearly dependent regression effects'' because the matrices $\textbf{X}$ and $\bm{\Phi}$ are allowed to be linearly dependent. Let $\bm{\epsilon}$ be an $n$-dimensional random vector representing measurement error with mean-zero, $\mathrm{cov}(\bm{\epsilon}) = \sigma^{2}\textbf{I}$, and $\sigma^{2}>0$. In this article, we assume $\bm{\epsilon}$ is Gaussian distributed; that is, $f_{SLMM}(\textbf{y}\vert \bm{\beta},\bm{\nu},\sigma^{2}) = N(\textbf{X}\bm{\beta}+\bm{\Phi}\bm{\nu},\sigma^{2}\textbf{I})$, where $f$ denotes a probability density function (pdf) and $N(\mu,\sigma^{2})$ is a shorthand for the normal distribution with mean $\mu$ and variance $\sigma^{2}>0$, and the subscript ``SLMM'' will indicate that the pdf appears in the expression of the hierarchical representation of the SLMM. 

Two key inferential questions that arise when considering the SLMM is the estimation of $\bm{\beta}$ and spatial prediction. In particular, if one splits the data vector $\textbf{y}$ into an observed data vector and a missing deta vector $\textbf{y} = (\textbf{y}_{o}^{\prime},\textbf{y}_{m}^{\prime})^{\prime}$ with $n_{o}$-dimensional $\textbf{y}_{o}$ observed, and $m$-dimensional $\textbf{y}_{m}$ missing, with $n = n_{o}+n_{m}$, one can use the SLMM to predict $\textbf{y}_{m}$ using $\textbf{y}_{o}$. In a similar manner one can use the SLMM to estimate $\bm{\beta}$ based on the observed data vector $\textbf{y}_{o}$ via the generalized least squares estimator. With any parametric model, one should be mindful of concerns with identifiability (i.e., a value of the parameter produces a unique value of the data model). In the context of Equation (\ref{eq:slmm}) it is well known that $\textbf{X}\bm{\beta}$ and $\bm{\Phi}\bm{\nu}$ are not necessarily identifiable \citep[e.g.,][for a discussion]{paciorek2010importance}. By non-identifiable we mean that two different sets of parameters produce the same value of the likelihood in (\ref{eq:slmm}), or similarly the data model in (\ref{eq:slmm}) when $\bm{\nu}$ is interpreted as a random effect. 

To address concerns with identifiability, some have emphasized that additional assumptions are needed  \citep{paciorek2010importance,hughes2017spatial}. That is, one can always reparameterize the SLMM in (\ref{eq:slmm}) as follows \citep[e.g. see][]{hanks2015restricted},
\begin{align}\label{eq:rsr}
	\textbf{y} &= \textbf{X}\bm{\delta} + (\textbf{I}-\textbf{P})\bm{\Phi}\bm{\nu} + \bm{\epsilon},
\end{align}
\noindent
where $\textbf{P} = \textbf{X}(\textbf{X}^{\prime}\textbf{X})^{-1}\textbf{X}^{\prime}$ and $\bm{\delta} = \bm{\beta} + (\textbf{X}^{\prime}\textbf{X})^{-1}\textbf{X}^{\prime}\bm{\Phi}\bm{\nu}$. Equation (\ref{eq:rsr}) is simply a reparameterization of the  SLMM and we use the word ``reparameterization,'' in the same context as \citet{hanks2015restricted}, where we have a mapping $\bm{\delta} = \bm{\beta} + (\textbf{X}^{\prime}\textbf{X})^{-1}\textbf{X}^{\prime}\bm{\Phi}\bm{\nu}$ that formally defines a change-of-variables. In general, we call $\bm{\delta}$ the ``orthogonalized regression effects,'' since $\textbf{X}$ and $(\textbf{I}-\textbf{P})$ are orthogonal to each other.

Restricted spatial regression (RSR) models have become a popular strategy in the literature, which simply add the assumption that $\bm{\delta} = \bm{\beta}$, which is often checked using the variance inflation factor (VIF) diagnostic. Key references include \citet{Reich} and \citet{hodges2010adding}. It is important to emphasize that RSR is not the process of orthogonalization presented in Equations (1) and (2) because it adds the assumption that $\bm{\beta}=\bm{\delta}$. An important motivation for the RSR is that $\textbf{X}\bm{\beta}$ and $\bm{\Phi}\bm{\nu}$ are not identifiable if there are no additional assumptions on $\bm{\beta}$ and $\bm{\delta}$ \citep{paciorek2010importance}, such as $\bm{\delta} = \bm{\beta}$. Identifiability of additive models has a rich history \citep{kotlarski1967characterizing}, and adding assumptions to the process of orthogonalization can be seen as one of many strategies to address this concern. This is particularly important to a classical interpretation of $\bm{\Phi}\bm{\nu}$ as a proxy for unmeasured covariates \citep{Clayton}. That is, if all the unmeasured confounders are spatially co-varying, and the spatial statistical model for $\bm{\Phi}\bm{\nu}$ accurately models this spatial dependence, then a spatially co-varying error term can account for unmeasured confounders. Under this perspective the identifiability issue discussed by \citet{paciorek2010importance} and others is important as the likelihood can not identify the latent mean $\textbf{X}\bm{\beta}$ and the ``unmeasured confounders'' $\bm{\Phi}\bm{\nu}$. 

 \citet{zimmerman2022deconfounding} and \citet{khan2022restricted} identified several serious inferential issues when the RSR is ``misspecified.'' That is, we say the RSR is misspecified when $\bm{\beta} \ne \bm{\delta}$, which implies that the SLMM is correctly specified. In particular, \citet{zimmerman2022deconfounding} made important discoveries that show when one incorrectly assumes $\bm{\beta} = \bm{\delta}$, and the data is generated according to the SLMM, RSR is generally inferior in terms of several types of inferences (i.e., in terms of variances of regression estimates, confidence intervals, and prediction error variances) in a frequentist context. \citet{khan2022restricted} made important insights on the posterior distribution for the precision parameter, and consequently, when one assumes $\bm{\beta} = \bm{\delta}$, Bayesian implementations of RSR produce posterior estimates of $\bm{\beta}$ {that are} equivalent to OLS with variances that go to zero as $n$ grows (i.e., an overspecified OLS estimate). 

The results from \citet{khan2022restricted} and \citet{zimmerman2022deconfounding} might lead someone to make, what we call, ``Conclusion 1.''\\


\noindent
{Conclusion 1: Inferences on the linearly dependent regression effects using the misspecified Bayesian RSR are ``generally inferior'' to that of the original SLMM.\\

\noindent	
In fact both works, arguably make several stronger conclusions, including convincing results that a  nonspatial model (i.e., $\bm{\Sigma}$ that is proportional to the identity) was preferable in terms of coverage of $\bm{\beta}$ than the misspecified RSR model in our Gaussian context. The RSR has historically been used for several types of inference in addition to regression estimation \citep[e.g., see][among others]{bradleyMSTM}, which is one reason why \citet{zimmerman2022deconfounding} investigate predictors derived from the misspecified RSR. In particular, \citet{zimmerman2022deconfounding} argue that a particular frequentist approach to RSR spatial prediction produces prediction error variances that lead to under-coverage. I am in complete agreement with Conclusion 1 and the additional criticisms outlined in these work. These works unequivocally show that using estimates from the traditional RSR to estimate a fixed value of $\bm{\beta}$ is the wrong thing to do if one wants to adjust for spatial confounding when one assumes $\bm{\beta}\ne \bm{\delta}$. In this article, we aim to discuss the RSR from a different perspective, that leads to different conclusions and interpretations.

Conclusion 1 is reached often when one expressly assumes the RSR is misspecified and the true data generating model is assumed to be the SLMM \citep{khan2023re}. However, in traditional model selection/comparison contexts one does not assume any candidate model is true or false to avoid confirmation bias \citep[e.g., see][for a standard reference on model selection]{burnham}. Then impartial selection criteria, such as the Bayes factor \citep{jeffreys1998theory} or strictly proper scoring rules \citep{gneiting2007strictly}, among others, might be used to compare and select models. In this article, we adopt a traditional model comparison perspective to avoid confirmation bias, and generally don't assume a data generating model that favors either the Bayesian SLMM or the misspecified Bayesian RSR. In fact, we show that the Bayes factor, strictly proper scoring rules, and out-of-sample predictive metrics between the SLMM and the misspecified RSR, under certain prior distributional assumptions, suggest that both models are equivalent. Additionally, since RSR misspecification implies $\bm{\beta} \ne \bm{\delta}$, we intend to compare equivalent estimands between the misspecified RSR and the SLMM. That is, we ask the question, ``how does the SLMM's posterior distribution of $\bm{\delta}$ compare to the posterior distribution of regression effects from the misspecified RSR?'' This leads us to the what we call ``Conclusion 2'' and ``Conclusion 3.''\\
	
	\noindent
	{Conclusion 2: Under certain prior specifications, posterior inferences on the orthogonal regression effects, hyperparameters, and missing values from the misspecified Bayesian RSR are equivalent to the inferences on the orthogonal regression effects, hyperparameters, and missing values using the original Bayesian SLMM. Additionally, when using these same prior specifications the marginal distributions of the data are equivalent between the misspecified Bayesian RSR and original Bayesian SLMM. }\\
	
\noindent
Specifically, when adopting the prior from \citet{Reich}, we find that the original SLMM's marginal posterior distribution for $\bm{\delta}$, hyperparameters, and $\textbf{y}_{m}$ is identical to the misspecified Bayesian RSR model's marginal posterior distribution, and that the marginal distribution of the data are identical between the two model. An immediate consequence of these equivalence relationships is that we see no difference among the aforementioned selection criteria. Thus, when adopting a formal model selection perspective we find several standard selection criteria view the original Bayesian SLMM and the misspecified Bayesian RSR as equivalent under certain (not all) prior specifications. This suggests that whenever the SLMM is reasonable, it would also be reasonable to use the misspecified RSR, under certain prior distributional assumptions, to determine the effect of a coviariate via selection criteria.

Conclusion 2 does not lead to any developments regarding estimation of linearly dependent regression effects from the misspecified RSR, which is at the core of the criticisms from the literature supporting Conclusion 1. In the misspecified setting (i.e., $\bm{\beta}\ne \bm{\delta}$), we suggest adjusting estimates from the traditional RSR to estimate linearly dependent regression effects using the posterior predictive strategy introduced by \citet{hanks2015restricted}. We show that this posterior predictive strategy can lead to an estimand that produces equivalent estimation of the original Bayesian SLMM's linearly dependent regression effects under certain prior specifications. This leads us to Conclusion 3.\\

	\noindent
{Conclusion 3: Under certain prior specifications, an augmented misspecified Bayesian RSR can reproduce the same inferences on linearly dependent regression effects (via a posterior predictive adjustment), missing values, spatial random effects, and hyperparameters as the original Bayesian SLMM. }\\

\noindent
Conclusions 2 and 3 both hold when using an improper prior on $\bm{\beta}$, which is the same prior used in \citet{Reich}.

New results are provided that show that one can sample directly from the augmented misspecified Bayesian RSR posterior distribution without the use of MCMC. The equivalence relationships described above show that this sampler can also be used to directly sample from the posterior distribution for the SLMM as well. Existing direct sampling strategies make use of the method of composition \citep{press2009subjective}, and often requires the use of a discrete uniform distribution to obtain closed form expressions of the marginal posterior distribution of hyperparameters \citep{zhang2023exact}. To our knowledge this is the first time the full posterior has been derived in closed form for Gaussian spatial linear mixed models that does not make use of discrete uniform prior distributions on hyperparameters. These new computational results combined with Conclusions 2 and 3 suggest that there is a clear practical reason to use the augmented misspecified Bayesian RSR (i.e., computation). Specifically, the augmented misspecified RSR produces the same posterior inferences as the SLMM (after appropriate adjustments), and is computationally efficient.

 We provide new insight on the use of a misspecified Bayesian RSR in the presence of unmeasured confounders. We explicitly show that posterior predictive summaries of $\bm{\delta}$ are invariant to unmeasured confounders and model misspecification of $\bm{\Phi}\bm{\nu}$. Ultimately, this is one of the original motivations for the RSR, however, we develop this property further via a semi-parametric representation. \citet{hanks2015restricted}'s adjustments to the misspecified RSR is then investigated in the context of unmeasured confounders. I interpret \citet{hanks2015restricted}'s posterior-predictive step to estimate $\bm{\beta}$ via the misspecified RSR as a type of transfer learning strategy \citep{weiss2016survey}, where a method-of-moments (MoM) approach is used to ensure the first moments match between the orthogonal and linearly dependent regression coefficients. Specifically, \citet{hanks2015restricted} makes use of a two step procedure, where one first implements a misspecified augmented RSR (i.e., a model that incorrectly assumes $\bm{\beta} = \bm{\delta}$), and then uses a posterior predictive step to adjust inferences to perform inference on linearly dependent regression effects. Such adjusted posterior predictive quantities are denoted with $\bm{\beta}_{trn}^{adj}$. We emphasize \citet{hanks2015restricted}'s two step procedure that we adopt for estimation of linearly dependent regression coefficients is not the traditional implementation of the RSR that restricts $\bm{\delta}=\bm{\beta}$. Instead we consider new restrictions that assume $\bm{\beta}_{trn}^{adj} = \bm{\beta}$ under various choices for the model for $\bm{\beta}_{trn}^{adj}$. In this article, \citet{hanks2015restricted}'s strategy is developed further by using $\bm{\delta}$ as data within a Bayesian hierarchical model that does not force the first moment of $\bm{\beta}$ and $\bm{\delta}$ to match. It is empirically shown in a simulation study that this transfer learning RSR strategy can improve on the SLMM, Spatial+ \citep{dupont2022spatial}, and double machine learning (DML) \citep{gilbert2021causal} for estimation of linearly dependent regression effects.
			
In Section 2 we demonstrate that Conclusions 2 and 3 hold for the traditional RSR from \citet{Reich}, and discuss the implication of the equivalence relationships that are found. Then in Section 3 and 4, two benefits of the RSR are introduced including robustness properties that motivate our new transfer learning approach, and a new sampling strategy that allows for independent replicates from the full posterior distributions (Section 4). Section 5 provides illustrations via simulations and an application to COVID-19 mortality data. We end with a discussion in Section 6.

			\vspace{-20pt}
			
			\section{Equivalence Relationships and their Implication on Selection Criteria} \label{sec:clarifications}

			Consider the following special case of the Bayesian SLMM written hierarchically as follows:
			\begin{align}
				\nonumber
				f_{SLMM}(\textbf{y}\vert  \bm{\beta},\bm{\nu},\sigma^{2})&= N(\textbf{X}\bm{\beta} + \bm{\nu},\sigma^{2}\textbf{I})\\
				\nonumber
				f_{SLMM}(\bm{\beta})  &= 1\\
				\nonumber
				f_{SLMM}(\bm{\nu}\vert \bm{\Sigma}_{\nu}) &= N(\bm{0}_{n},\bm{\Sigma}_{\nu})\\
				\nonumber
				&f(\bm{\Sigma}_{\nu})\\
				\label{eq:slmm:bhm2}
				&f(\sigma^{2}).
			\end{align}
			\noindent
			Upon multiplying each level in the Bayesian hierarchical model we obtain the joint distribution,
			\begin{equation*}
				f_{SLMM}(\textbf{y},\bm{\beta},\bm{\nu},\bm{\Sigma}_{\nu},\sigma^{2}) = f_{SLMM}(\textbf{y}\vert  \bm{\beta},\bm{\nu},\sigma^{2})	f_{SLMM}(\bm{\nu}\vert \bm{\Sigma}_{\nu})f(\bm{\Sigma}_{\nu})f(\sigma^{2}).
			\end{equation*}
			This is a special case of the Bayesian SLMM model that sets $\bm{\Phi}=\textbf{I}$, assumes an improper prior for $\bm{\beta}$, and $\bm{\Sigma}=\bm{\Sigma}_{\nu}$. Common choices for parametric structures for $\bm{\Sigma}_{\nu}$ include basis function expansions \citep{cressie-wikle-book}, the covariance matrix from a conditional autoregressive model \citep{besagYorkMollie}, or covariance functions from geostatistical model such as the Mat\'{e}rn covariogram \citep{banerjee-etal-2004}. This model has the important property that $\bm{\beta} \ne \bm{\beta} + (\textbf{X}^{\prime}\textbf{X})^{-1}\textbf{X}^{\prime}\bm{\nu}= \bm{\delta}$ almost surely, and is hence, correctly specified. 
			
			Throughout Sections~2.1$\--$2.4 we will assume the data is generated according to the model in (\ref{eq:slmm:bhm2}). That is, a classical Bayesian perspective implies that the data generating mechanism is equal to the marginal distribution of the data \citep{walker2013bayesian}, e.g., see discussions surrounding the Type II maximum likelihood \citep{lehmann1998theory}, Bayes factors, and  De Finetti's representation theorem \citep[e.g., see][for a standard reference]{schervish2012theory}. Specifically, Bayesian hierarchical models defines a distribution for the data via the marginal distribution,
			\begin{equation*}
				\mathrm{\textbf{Generating}}\hspace{3pt}\mathrm{\textbf{Model}}:\hspace{2pt}f_{SLMM}(\textbf{y}) = \int \int \int \int f_{SLMM}(\textbf{y},\bm{\beta},\bm{\nu},\bm{\Sigma}_{\nu},\sigma^{2})\hspace{2pt} d\bm{\beta}\hspace{2pt}d\bm{\nu}\hspace{2pt}d\bm{\Sigma}_{\nu}\hspace{2pt}d\sigma^{2}.
			\end{equation*}
			We call this the generative model because if $\textbf{y}^{*}$ is simulated from $f_{SLMM}(\textbf{y})$ it is equal in distribution to $\textbf{y}$ assuming (\ref{eq:slmm:bhm2}). This data generating mechanism above assumes both $\bm{\beta}$ and $\bm{\nu}$ are random vectors, and is considered as a data generating mechanism that is impartial (see Section 2.5) when comparing RSR to the SLMM. This choice, however, effects the interpretation of $\bm{\beta}$. In this article, when we say a random quantity has ``physical meaning'' we mean that it can be interpreted as a ``random effect,'' using the traditional nomenclature of fixed, random, and mixed effects. The vector $\bm{\nu}$ is traditionally interpreted as a random effect with elements that might arise from a random process (e.g., see Cressie and Wikle, 2011, for a discussion of the ``data model, process model, and parameter model'' representation of a hierarchical model). However, $\bm{\beta}$ is traditionally not considered a random effect. Thus, under this data generating mechanism, $\bm{\beta}$ is \textit{not a physically meaningful random quantity} because it is a random vector that is not a random effect. Inferences on $\bm{\beta}$ are still of interest because one way to generate predictive data is to first simulate a value $\bm{\beta}^{*}$ from  $f_{SLMM}(\bm{\beta}\vert \textbf{y}_{o})$ and then subsequently simulate from $f_{SLMM}(\textbf{y}_{m}\vert \bm{\beta}^{*})$. That is, the value of the non-physical construct $\bm{\beta}^{*}$ is paired with the value of something that can be physically realized via random effects, namely, predictive data (e.g., if an element of $\bm{\beta}^{*}$ is zero then the corresponding column in $\textbf{X}$ is not useful for summarizing the conditional mean of posterior predictive data). Or equivalently, one might generate posterior predictive data by first sampling $\bm{\delta}^{*}$ from  $f_{SLMM}(\bm{\delta}\vert \textbf{y}_{o})$ and then subsequently simulate from $f_{SLMM}(\textbf{y}_{m}\vert \bm{\delta}^{*})$. Similarly, $\bm{\delta}$ can be interpreted as a parameter defining the conditional mean for $\textbf{y}_{m}$. The appearance of the random vector $\bm{\beta}$ in the definition of $\bm{\delta}$ (i.e.,
			$\bm{\delta} = \bm{\beta} + (\mathbf{X}'\mathbf{X})^{-1}\mathbf{X}'\bm{\nu}$)
			suggests that $\bm{\delta}$ likewise lacks a non-physical interpretation.
			Under our assumed data-generating mechanism, $\bm{\beta}$ is not a
			physically meaningful quantity  (only mathematically meaningful as a conditional mean), and consequently neither is
			$\bm{\delta}$, since it is defined in terms of $\bm{\beta}$.
			
			In the frequentist setting, we have $\textbf{y} = \textbf{u} + \bm{\epsilon}$ where it is assumed $\textbf{u}\sim N(\textbf{X}\bm{\beta}_{0}, \textbf{B}\bm{\Sigma}_{0}\textbf{B}^{\prime})$ and $\bm{\epsilon}\sim N(\bm{0}_{n},\sigma_{0}^{2})$, with parameters $\bm{\beta}_{0}$ (referred to as a ``fixed effect'' in this setting), $\bm{\Sigma}_{0}$, and $\sigma_{0}^{2}$ are well-defined fixed and unknown values, $\textbf{u}$ is a random effect vector, and $\bm{\epsilon}$ is a random effect vector often interpreted as measurement error. The parameters $\bm{\beta}_{0}$, $\bm{\Sigma}_{0}$, and $\sigma_{0}^{2}$ are interpretable as they they define the model for random effects (recall we call random quantities that are physically meaningful ``random effects''). In particular, $\bm{\beta}_{0}$ and $\bm{\Sigma}_{0}$ defines the mean and covariance matrix for the random effect vector $\textbf{u}$. In this setting $\bm{\delta}_{0} = \bm{\beta}_{0}+(\textbf{X}^{\prime}\textbf{X})^{-1}\textbf{X}^{\prime}(\textbf{u}-\textbf{X}\bm{\beta}_{0})$ becomes difficult to interpret as it is conditionally biased (conditioned on $\textbf{u}-\textbf{X}\bm{\beta}_{0}$) for the free parameter $\bm{\beta}_{0}$ that defines the conditional mean for the random effect vector $\textbf{u}$. Thus, in the frequentist paradigm linearly dependent regression effects are interpretable while orthogonalized regression effects are conditionally biased for an interpretable quantity. However, under our assumed data generating mechanism both quantities $\bm{\beta}$ and $\bm{\delta}$ are random, both can be used to describe the conditional mean of posterior predictive data, and both are not physically interpretable. Consequently, one might equally consider estimating both quantities under our assumed data generating mechanism to summarize the conditional mean of predictive data, which is not the case when assuming the frequentist paradigm.
			
			One can differ from our data generating mechanism by assuming a ``true'' value of $\bm{\beta}$, and still fit the Bayesian SLMM stated above. In a subjective Bayesian context one might specify their prior beliefs about the value of $\bm{\beta}$, and of course, our beliefs may not be reflective of how the data is generated, which can introduce bias in the results. In our context, that would imply that the prior distribution for $\bm{\beta}$ is misspecified (e.g., the data is generated assuming the improper prior on $\bm{\beta}$, and we fit a Bayesian hierarchical model with a Gaussian prior). For those new to this paradigm, see the nonparametric empirical Bayesian literature that aims to estimate the ``true prior distribution'' \citep[e.g., see][for a recent article]{banerjee2020nonparametric}. Thus, one can still assume that there is a true fixed value $\bm{\beta}_{0}$ (a ``fixed effect'') and apply a Bayesian hierarchical model with an incorrectly specified prior distribution on $\bm{\beta}$. In this case, the true prior distribution for $\bm{\beta}$ is a point mass on $\bm{\beta}_{0}$, and the Bayesian SLMM as defined in this article would incorrectly use an improper prior distribution. To address prior misspecification, when assuming $\bm{\beta}$ has a point mass on $\bm{\beta}_{0}$, posterior consistency results are often used to show that the posterior distribution concentrates on $\bm{\beta}_{0}$ as $n$ grows despite the prior misspecification \citep{yang2015semiparametric}. Additionally, selection criteria can be used to select models indexed by different prior distributional assumptions. In Section 2.1$\--$2.4, we consider the setting where the SLMM is correctly specified with its improper prior in its full hierarchy. In Sections 2.5 and 3, we consider other cases where the SLMM and RSR are both misspecified in various ways. Specifically, in Section 2.5 we consider the case when the data generating mechanism is unknown and not necessarily equal any of our candidate models, and in Section 3 we assume a hierarchical model with certain levels of the hierarchy unknown.

			\subsection{Misspecified Bayesian RSR Models}\label{sec:Eq3}

Suppose we are unaware that the data is distributed according to the density $f_{SLMM}(\textbf{y})$ derived from (\ref{eq:slmm:bhm2}), and we assume the RSR model from \citet{Reich} given by:
\begin{align}
	\nonumber
	f_{RSR}(\textbf{y}\vert \bm{\beta}_{RSR},\bm{\nu}_{RSR},\sigma^{2})&= N(\textbf{X}\bm{\beta}_{RSR} + \textbf{L}\bm{\nu}_{RSR},\sigma^{2}\textbf{I})\\
	\nonumber
	f_{RSR}(\bm{\beta}_{RSR})  &= 1\\
	\nonumber
	f_{RSR}(\bm{\nu}_{RSR}\vert \bm{\Sigma}_{\nu}) &= N(\bm{0}_{n},\textbf{L}^{\prime}\bm{\Sigma}_{\nu}\textbf{L})\\
	\nonumber
	&f(\bm{\Sigma}_{\nu})\\
	\label{eq:rsr:bhm}
	&f(\sigma^{2}),
\end{align}
\noindent 
where $\textbf{L}$ is the $n\times (n-p)$ eigenvectors of $\textbf{I}-\textbf{P}$ such that $\textbf{L}\textbf{L}^{\prime} = \textbf{I}-\textbf{P}$. Of course, the RSR itself is a special case of the SLMM in (\ref{eq:slmm}) when $r = n-p$, $\bm{\Phi} = \textbf{L}$, and $\bm{\Sigma} = \textbf{L}^{\prime}\bm{\Sigma}_{\nu}\textbf{L}$. If one considers the reparameterization in (\ref{eq:rsr}) for this special case of the SLMM, we have $\bm{\delta}_{RSR} = \bm{\beta}_{RSR} + (\textbf{X}^{\prime}\textbf{X})^{-1}\textbf{X}^{\prime}\textbf{L}\bm{\nu}_{RSR} = \bm{\beta}_{RSR}$, and is consequently, misspecified. Thus, for this misspecified Bayesian RSR, it is correct to call $\bm{\beta}_{RSR}$ the orthogonal regression effects, since $\bm{\beta}_{RSR}$ is equal to the orthogonal regression effects $\bm{\delta}_{RSR}$. 

The primary motivation for this misspecified Bayesian RSR is that when one derives the predictive distribution for $\bm{\beta}_{RSR}$ we obtain \citep[e.g., see][among others]{Reich,khan2022restricted}\\
\begin{equation}\label{eq:ols}
	\bm{\beta}_{RSR}\vert \textbf{y},\sigma^{2}\sim N\left\lbrace(\textbf{X}^{\prime}\textbf{X})^{-1}\textbf{X}^{\prime}\textbf{y},\sigma^{2}(\textbf{X}^{\prime}\textbf{X})^{-1}\right\rbrace.
\end{equation}
\noindent
 When using the term ``predictive distribution'' for regression effects we mean the conditional distribution of the regression effect given the data, covariates, and covariance parameters. The predictive distribution for $\bm{\beta}_{RSR}$ centers on the ordinary least squares (OLS) estimator. The OLS estimator does not make use of spatial correlations, and is an conditionally unbiased estimator for $\bm{\beta}$ assuming the SLMM and no unmeasured confounders. This property is attractive in the setting when $\bm{\Sigma}_{\nu}$ is possibly misspecified by the presence of unmeasured covariates, so that the error induced by misspecification is avoided.

The joint pdf of all parameters and random effects from the RSR is given by,
\begin{equation}
	\label{eq:rsr:joint2}
	f_{RSR}(\textbf{y},\bm{\beta}_{RSR},\bm{\nu}_{RSR},\bm{\Sigma}_{\nu},\sigma^{2})= f_{RSR}(\textbf{y}\vert  \bm{\beta}_{RSR},\bm{\nu}_{RSR},\sigma^{2})f_{RSR}(\bm{\nu}_{RSR}\vert \bm{\Sigma}_{\nu},\sigma^{2})f(\bm{\Sigma}_{\nu})f(\sigma^{2}).
\end{equation}
\noindent
This expression of the RSR model is somewhat different from our exposition in the Introduction, where the coefficients of the spatial random effects were $(\textbf{I}-\textbf{P})$ instead of coefficients for $\textbf{L}$  in Equation~(2). However, it is easy to verify that $\textbf{L}\bm{\nu}_{RSR}$ is equal in distribution to $(\textbf{I}-\textbf{P})\bm{\nu}$, which leads to the following alternative ``augmented RSR'' :
\begin{align}
	\nonumber
	f_{aRSR}(\textbf{y}\vert  \bm{\beta}_{RSR},\bm{\nu},\sigma^{2})&= N(\textbf{X}\bm{\beta}_{RSR} + (\textbf{I}-\textbf{P})\bm{\nu},\sigma^{2}\textbf{I})\\
	\nonumber
	f_{RSR}(\bm{\beta}_{RSR})  &= 1\\
	\nonumber
	f_{SLMM}(\bm{\nu}\vert \bm{\Sigma}_{\nu}) &= N(\bm{0}_{n},\bm{\Sigma}_{\nu})\\
	\nonumber
	&f(\bm{\Sigma}_{\nu})\\
	\label{eq:rsr:bhm:aug}
	&f(\sigma^{2}),
\end{align}
\noindent 
where ``aRSR'' should be read as ``augmented RSR,'' and is the model considered in \citet{hanks2015restricted}. This leads to the following joint distribution
\begin{equation}
	\label{eq:rsr:joint}
	f_{aRSR}(\textbf{y},\bm{\beta}_{RSR},{\bm{\nu}},\bm{\Sigma}_{\nu},\sigma^{2})= f_{aRSR}(\textbf{y}\vert  \bm{\beta}_{RSR},{\bm{\nu}},\sigma^{2})f_{SLMM}({\bm{\nu}}\vert \bm{\Sigma}_{\nu},\sigma^{2})f(\bm{\Sigma}_{\nu})f(\sigma^{2}).
\end{equation}
This augmented model is also a special case of the SLMM in (\ref{eq:slmm}) when $r = n$, $\bm{\Phi} = \textbf{I}-\textbf{P}$, and $\bm{\Sigma} = \bm{\Sigma}_{\nu}$. Again if one considers the reparameterization in (\ref{eq:rsr}) for this SLMM, we have $\bm{\delta}_{RSR} = \bm{\beta}_{RSR} + (\textbf{X}^{\prime}\textbf{X})^{-1}\textbf{X}^{\prime}(\textbf{I}-\textbf{P}){\bm{\nu}} = \bm{\beta}_{RSR}$, and is also misspecified. The two Bayesian hierarchical models in (\ref{eq:rsr:bhm}) and (\ref{eq:rsr:bhm:aug}) are both considered two different data augmentation strategies because they both reproduce the same likelihood as follows,
\begin{align}
	\nonumber
	\int f_{RSR}(\textbf{y}\vert \textbf{X}, \bm{\beta}_{RSR},\bm{\nu}_{RSR},\sigma^{2}) f(\bm{\nu}_{RSR}\vert \bm{\Sigma}_{\nu}) d\bm{\nu}_{RSR} &= \int f_{aRSR}(\textbf{y}\vert \textbf{X}, \bm{\beta}_{RSR},{\bm{\nu}},\sigma^{2}) f_{SLMM}({\bm{\nu}}\vert \bm{\Sigma}_{\nu},\sigma^{2}) d{\bm{\nu}}\\
	\label{eq:rsr:aug}
	& = N\left\lbrace\textbf{X}\bm{\beta}_{RSR},(\textbf{I}-\textbf{P})\bm{\Sigma}_{\nu}(\textbf{I}-\textbf{P}))\right\rbrace,
\end{align}
\noindent
which implies
\begin{equation}\label{eq:aug:result}
	f_{aRSR}(\textbf{y},\bm{\beta}_{RSR},\bm{\Sigma}_{\nu},\sigma^{2}) = f_{RSR}(\textbf{y},\bm{\beta}_{RSR},\bm{\Sigma}_{\nu},\sigma^{2}),
\end{equation}
\noindent
so that,
\begin{align*}
	&f_{aRSR}(\bm{\beta}_{RSR},\bm{\Sigma}_{\nu},\sigma^{2},\textbf{y}_{m}\vert \textbf{y}_{o}) = f_{RSR}(\bm{\beta}_{RSR},\bm{\Sigma}_{\nu},\sigma^{2},\textbf{y}_{m}\vert \textbf{y}_{o}).
\end{align*} 
Thus, aRSR in (\ref{eq:rsr:bhm:aug}) and RSR in (\ref{eq:rsr:bhm}) produce the same posterior summaries on $\bm{\beta}_{RSR}$, hyperparameters, and predictions of $\textbf{y}_{m}$. As we will see, considering both augmented models in (\ref{eq:rsr:bhm}) and (\ref{eq:rsr:bhm:aug}) will play a crucial role for arriving to Conclusions 2 and 3. In particular, Equation (\ref{eq:rsr:bhm}) is more consistent with the traditional RSR, while Equation (\ref{eq:rsr:bhm:aug}) leads to \citep{hanks2015restricted}'s treatment of the RSR. Equation (\ref{eq:rsr:bhm:aug}) is preferable, as it allows one to perform a posterior predictive adjustments to recover the linearly dependent regression coefficients under suitable prior assumptions, where Equation (\ref{eq:rsr:bhm}) does not allow one to do this.

\subsection{Demonstrating Conclusion 2}\label{sec:Conclusion 2} Suppose we assume the data and parameters are generated according to the SLMM\\ $f_{SLMM}(\textbf{y},\bm{\beta},\bm{\nu},\bm{\Sigma}_{\nu},\sigma^{2})$. Now consider applying a formal change-of-variables to \\
 $f_{SLMM}(\textbf{y},\bm{\beta},\bm{\nu},\bm{\Sigma}_{\nu},\sigma^{2})$ via the mapping 
\begin{equation}
	\left(\begin{array}{c}
		\bm{\delta}\\
		\widetilde{\bm{\nu}}
	\end{array}\right) = 	\left(\begin{array}{c}
		\bm{\beta}+ (\textbf{X}^{\prime}\textbf{X})^{-1}\textbf{X}^{\prime}\bm{\nu}\\
		\bm{\nu}
	\end{array}\right)
	\label{eq:changeofvariables}
\end{equation} 
\noindent
with Jacobian
\begin{equation*}
	det\left(\begin{array}{cc}
		\textbf{I}_{p}& -(\textbf{X}^{\prime}\textbf{X})^{-1}\textbf{X}^{\prime}\\
		\bm{0}_{n,p}& \textbf{I}
	\end{array}\right) = 1,
\end{equation*}
where $\textbf{I}_{p}$ is a $p\times p$ identity matrix, $\bm{0}_{n,p}$ is a $n\times p$ matrix of zeros. Standard change-of-variables involves substituting the inverse transformation and multiplying by the Jacobian to obtain:
\begin{align}
	\nonumber
	&f_{SLMM}(\textbf{y},\bm{\delta},\widetilde{\bm{\nu}},\bm{\Sigma}_{\nu},\sigma^{2})\\
	\nonumber
	& = f_{SLMM}(\textbf{y},\bm{\beta} = \bm{\delta}-(\textbf{X}^{\prime}\textbf{X})^{-1}\textbf{X}^{\prime}\widetilde{\bm{\nu}} ,\widetilde{\bm{\nu}},\bm{\Sigma}_{\nu},\sigma^{2})\\
	\nonumber
	&= f_{SLMM}(\textbf{y}\vert \bm{\beta} = \bm{\delta}-(\textbf{X}^{\prime}\textbf{X})^{-1}\textbf{X}^{\prime}\widetilde{\bm{\nu}},\widetilde{\bm{\nu}},\sigma^{2})f_{SLMM}(\widetilde{\bm{\nu}}\vert \bm{\Sigma}_{\nu})f(\bm{\Sigma}_{\nu})f(\sigma^{2})\\
	\nonumber
	& = f_{aRSR}(\textbf{y}\vert  \bm{\delta},\widetilde{\bm{\nu}},\sigma^{2})f_{SLMM}(\widetilde{\bm{\nu}}\vert \bm{\Sigma}_{\nu})f(\bm{\Sigma}_{\nu})f(\sigma^{2})\\
	\label{eq:slmm:torsr2}
	&= f_{aRSR}(\textbf{y},\bm{\delta},\widetilde{\bm{\nu}},\bm{\Sigma}_{\nu},\sigma^{2})
\end{align}

which follows immediately from the well-known fact that the likelihood is unidentifiable \citep{paciorek2010importance,hanks2015restricted},
\begin{equation}
	f_{SLMM}(\textbf{y}\vert \textbf{X}, \bm{\beta} = \bm{\delta}-(\textbf{X}^{\prime}\textbf{X})^{-1}\textbf{X}^{\prime}\bm{\nu},\bm{\nu},\sigma^{2}) = N(\textbf{X}\bm{\delta} + (\textbf{I}-\textbf{P})\bm{\nu},\sigma^{2}\textbf{I}) = f_{aRSR}(\textbf{y}\vert \textbf{X}, \bm{\delta},\bm{\nu},\sigma^{2}),
\end{equation}
and from the fact that that the prior for $\bm{\beta}$ is unchanged since $f(\bm{\beta} = \bm{\delta} - (\textbf{X}^{\prime}\textbf{X})^{-1}\textbf{X}^{\prime}\bm{\nu}) = f(\bm{\beta}) = 1$. In fact, throughout Section~\ref{sec:clarifications} one can substitute $f(\bm{\beta}) = 1$ with any location invariant prior distribution (see Section 2.4). When integrating across $\widetilde{\bm{\nu}}$ in Equation (\ref{eq:slmm:torsr2}), it follows from (\ref{eq:aug:result}), that
\begin{equation*}
f_{SLMM}(\textbf{y},\bm{\delta},\bm{\Sigma}_{\nu},\sigma^{2}) = f_{RSR}(\textbf{y},\bm{\delta},\bm{\Sigma}_{\nu},\sigma^{2}).
\end{equation*}
\noindent
This implies,

\begin{align}
	\nonumber
	&f_{SLMM}(\textbf{y}_{o}) =  f_{RSR}(\textbf{y}_{o}) = f_{aRSR}(\textbf{y}_{o})\\
\label{eq:orthoequaltoslmm}
&f_{SLMM}(\bm{\delta},\bm{\Sigma}_{\nu},\sigma^{2},\textbf{y}_{m}\vert \textbf{y}_{o}) = f_{RSR}(\bm{\delta},\bm{\Sigma}_{\nu},\sigma^{2},\textbf{y}_{m}\vert \textbf{y}_{o})= f_{aRSR}(\bm{\delta},\bm{\Sigma}_{\nu},\sigma^{2},\textbf{y}_{m}\vert \textbf{y}_{o}),
\end{align}
\noindent
which verifies Conclusion 2. Thus, all posterior inference on the orthogonal regression coefficients, hyperparameters, and $\textbf{y}_{m}$ (including predictions via posterior means and variances) are identical between the misspecified Bayesian RSR and the correctly specified SLMM. Additionally, the data generating mechanism (i.e., the model assumed for the data, $f_{SLMM}(\textbf{y}_{o})$) are identical between the SLMM, RSR, and aRSR.

\subsection{Demonstrating Conclusion 3}\label{sec:Conclusion 3}

 Consider the following change-of-variables for the misspecified augmented RSR model $f_{aRSR}(\textbf{y},\bm{\beta}_{RSR},\bm{\nu},\bm{\Sigma}_{\nu},\sigma^{2})$ based on the mapping,
\begin{equation}\label{eq:c-o-v}
	\left(\begin{array}{c}
		\bm{\beta}^{adj}\\
		\widetilde{\bm{\nu}}
	\end{array}\right) = 	\left(\begin{array}{c}
		\bm{\beta}_{RSR}- (\textbf{X}^{\prime}\textbf{X})^{-1}\textbf{X}^{\prime}\bm{\nu}\\
		\bm{\nu}
	\end{array}\right) 
\end{equation} 
\noindent
which has Jacobian
\begin{equation*}
	det\left(\begin{array}{cc}
		\textbf{I}_{p}& (\textbf{X}^{\prime}\textbf{X})^{-1}\textbf{X}^{\prime}\\
		\bm{0}_{n,p}& \textbf{I}
	\end{array}\right) = 1.
\end{equation*}

\noindent
At this stage we would like to remind the reader that the aRSR is misspecified as it assumes linearly dependent regression effects are equivalent to orthogonal regression effects (i.e., $\bm{\beta}_{RSR} = \bm{\delta}_{RSR}$). Moreover, the transformation that defines $\bm{\beta}^{adj}$ is interpreted as a posterior predictive step needed to adjust for misspecification. Similar to Equation (\ref{eq:slmm:torsr2}), we arrive at
\begin{align}
	\label{eq:rsr:toslmm}
	f_{aRSR}(\textbf{y},\bm{\beta}^{adj},\widetilde{\bm{\nu}},\bm{\Sigma}_{\nu},\sigma^{2}) = f_{SLMM}(\textbf{y}\vert \bm{\beta}^{adj},\widetilde{\bm{\nu}},\sigma^{2})f_{SLMM}(\bm{\nu}\vert \bm{\Sigma}_{\nu})f(\bm{\Sigma}_{\nu})f(\sigma^{2}).
\end{align}
It is important to recognize that Equation (\ref{eq:rsr:toslmm}) reproduces the Bayesian hierarchical model for an SLMM. Specifically, Equation (\ref{eq:rsr:toslmm}) can be re-expressed as,
\begin{align}
	\nonumber 
	f_{SLMM}(\textbf{y}\vert \bm{\beta}^{adj},\widetilde{\bm{\nu}},\sigma^{2}) & = N(\textbf{X}\bm{\beta}^{adj}+\widetilde{\bm{\nu}},\sigma^{2}\textbf{I})\\
	\nonumber
	f_{SLMM}(\bm{\beta}^{adj})  &= 1\\
	\nonumber
	f_{SLMM}(\widetilde{\bm{\nu}}\vert \bm{\Sigma}_{\nu}) &= N(\bm{0}_{n},\bm{\Sigma}_{\nu})\\
	\nonumber
	&f(\bm{\Sigma}_{\nu})\\
	\label{eq:slmm:bhm}
	&f(\sigma^{2})
\end{align}
\noindent
and notice that the model in (\ref{eq:slmm:bhm}) is an SLMM with improper prior on $\bm{\beta}^{adj}$. Thus, one can interpret $\bm{\beta}^{adj}$ as linearly dependent regression effects. When using a misspecified augmented RSR (i.e., $\bm{\beta}_{RSR}= \bm{\delta}_{RSR}$), $\bm{\beta}_{RSR}$ is interpreted as the ``orthogonalized regression effects'' and $\bm{\beta}^{adj}$ is interpreted as the ``linearly dependent regression effects.''

It follows from Equation (\ref{eq:rsr:toslmm}) that,

\begin{equation}\label{eq:rsrequalsslmm}
f_{aRSR}(\bm{\beta}^{adj},\widetilde{\bm{\nu}},\bm{\Sigma}_{\nu},\sigma^{2},\textbf{y}_{m}\vert \textbf{y}_{o})=f_{SLMM}(\bm{\beta}^{adj},\widetilde{\bm{\nu}},\bm{\Sigma}_{\nu},\sigma^{2},\textbf{y}_{m}\vert \textbf{y}_{o}),
\end{equation}
\noindent
which verifies Conclusion 3. Thus, the misspecified aRSR can be used to produce identical inferences on the linearly dependent regression coefficients as the correctly specified SLMM.

\begin{figure}[t!]
	\begin{center}
		\includegraphics[width=4in]{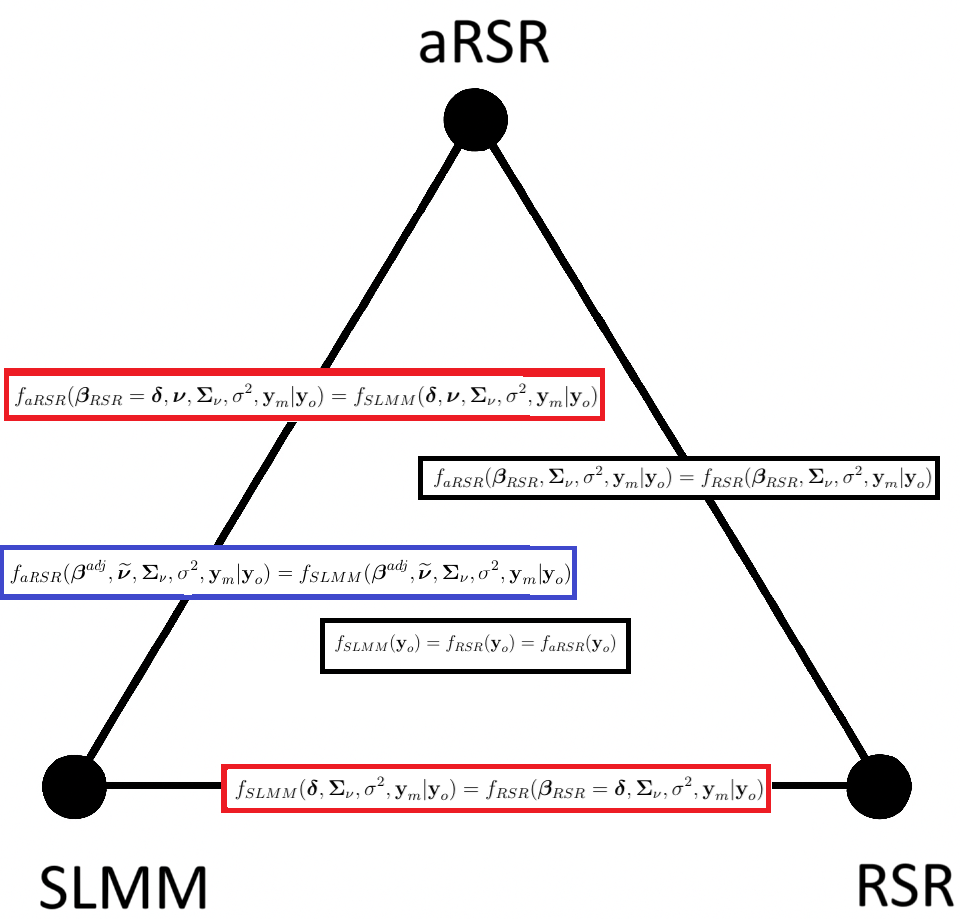}
	\end{center}
	\caption{A diagram of equivalence relationships between Bayesian aRSR, SLMM, RSR with improper priors on $\bm{\beta}$ and $\bm{\beta}_{RSR}$. The nodes are labeled by a density from one of the three models. The equations overlapping the edge between two nodes indicates an equivalence relationship between the models indicated by the node. The two equations outlined by red start with the SLMM and apply the change-of-variables $\bm{\delta} = \bm{\beta} + (\textbf{X}^{\prime}\textbf{X})^{-1}\textbf{X}^{\prime}\bm{\nu}$.  The equation outlined by blue start with the aRSR and apply the change-of-variables $\bm{\beta}^{adj} = \bm{\beta}_{RSR} - (\textbf{X}^{\prime}\textbf{X})^{-1}\textbf{X}^{\prime}\bm{\nu}$. The equations outlined in black follow from data augmentation. }\label{fig:first}
\end{figure}
\subsection{The Role of Identifiability and the use of Improper Prior Distributions} In Figure 1, we outline the equivalence relationships that are present between Bayesian RSR, aRSR, and SLMM models defined in Section~\ref{sec:clarifications}. We see that the misspecified Bayesian RSR and aRSR model's posterior inference on $\bm{\beta}_{RSR}$ and hyperparameters are equivalent to the original Bayesian SLMM's posterior inference on $\bm{\delta}$ and hyperparameters. Similarly, the aRSR model's posterior inference on $\bm{\beta}^{adj}$, $\widetilde{\bm{\nu}}$, and hyperparameters is equivalent to that of the orginal Bayesian SLMM's posterior inference on $\bm{\beta}$, $\bm{\nu}$, and hyperparameters. Additionally, posterior inference on missing values are equivalent between all three models (i.e., aRSR, RSR, and SLMM). These equivalence relationships arise for two main reasons: identifiability and the use of a location invariant prior distribution for $\bm{\beta}$. By location invariant (LI) we mean $f_{SLMM}(\bm{\beta} = \textbf{b}) = f_{SLMM}(\bm{\beta} = \textbf{b}+\textbf{c})$ for any real $\textbf{b}$ and $\textbf{c}$, which holds for $f_{SLMM}(\bm{\beta})\equiv 1$. \\

\noindent
\textbf{Proposition 1:} Consider the model SLMM model in (\ref{eq:slmm:bhm2}), the RSR model in (\ref{eq:rsr:bhm}), and the aRSR in (\ref{eq:rsr:bhm:aug}), and replace the improper prior distributions $f_{SLMM}(\bm{\beta}) = 1$ and $f_{RSR}(\bm{\beta}) = 1$ with any location invariant prior $f_{SLMM}^{(LI)}(\bm{\beta})$ and $f_{RSR}^{(LI)}(\bm{\beta}_{RSR})$. Then,
\begin{align*} 
&f_{SLMM}(\textbf{y}_{o}) =  f_{RSR}(\textbf{y}_{o}) = f_{aRSR}(\textbf{y}_{o})\\
 &f_{SLMM}(\bm{\delta},\bm{\Sigma}_{\nu},\sigma^{2},\textbf{y}_{m}\vert \textbf{y}_{o}) = f_{RSR}(\bm{\delta},\bm{\Sigma}_{\nu},\sigma^{2},\textbf{y}_{m}\vert \textbf{y}_{o})= f_{aRSR}(\bm{\delta},\bm{\Sigma}_{\nu},\sigma^{2},\textbf{y}_{m}\vert \textbf{y}_{o})\\
&f_{aRSR}(\bm{\beta}^{adj},\widetilde{\bm{\nu}},\bm{\Sigma}_{\nu},\sigma^{2},\textbf{y}_{m}\vert \textbf{y}_{o})=f_{SLMM}(\bm{\beta}^{adj},\widetilde{\bm{\nu}},\bm{\Sigma}_{\nu},\sigma^{2},\textbf{y}_{m}\vert \textbf{y}_{o}).
\end{align*}
\noindent
 \textbf{Proof:} The proof that$f_{SLMM}(\textbf{y}_{o}) =  f_{RSR}(\textbf{y}_{o}) = f_{aRSR}(\textbf{y}_{o})$ and $f_{SLMM}(\bm{\delta},\bm{\Sigma}_{\nu},\sigma^{2},\textbf{y}_{m}\vert \textbf{y}_{o}) = f_{RSR}(\bm{\delta},\bm{\Sigma}_{\nu},\sigma^{2},\textbf{y}_{m}\vert \textbf{y}_{o})= f_{aRSR}(\bm{\delta},\bm{\Sigma}_{\nu},\sigma^{2},\textbf{y}_{m}\vert \textbf{y}_{o})$ follows the same steps as Equations (\ref{eq:changeofvariables})$\--$ (\ref{eq:orthoequaltoslmm}), with the minor change of multiplying by $f_{SLMM}^{(LI)}(\bm{\beta} = \bm{\delta} - (\textbf{X}^{\prime}\textbf{X})^{-1}\textbf{x}^{\prime}\widetilde{\bm{\nu}}) = f_{SLMM}^{(LI)}(\bm{\beta} = \bm{\delta})$ in Equation (\ref{eq:slmm:torsr2}). Similarly, $f_{aRSR}(\bm{\beta}^{adj},\widetilde{\bm{\nu}},\bm{\Sigma}_{\nu},\sigma^{2},\textbf{y}_{m}\vert \textbf{y}_{o})=f_{SLMM}(\bm{\beta}^{adj},\widetilde{\bm{\nu}},\bm{\Sigma}_{\nu},\sigma^{2},\textbf{y}_{m}\vert \textbf{y}_{o})$ follows follows the same steps as Equations (\ref{eq:c-o-v})$\--$ (\ref{eq:rsrequalsslmm}), with the minor change of multiplying by $f_{RSR}^{(LI)}(\bm{\beta}^{adj} = \bm{\beta}_{RSR} - (\textbf{X}^{\prime}\textbf{X})^{-1}\textbf{x}^{\prime}\widetilde{\bm{\nu}}) = f_{RSR}^{(LI)}(\bm{\beta}^{adj} = \bm{\beta}_{RSR}) = f_{SLMM}^{(LI)}(\bm{\beta}^{adj} = \bm{\beta}_{RSR})$ in Equation (\ref{eq:rsr:toslmm}).

We emphasize that these equivalence relationships offer important clarifications to the literature. A recurring discussion in the literature is that there is a difference between the ``data generating'' model (i.e., SLMM) and ``the analysis model'' (i.e., aRSR) \citep{khan2022restricted,gilbert2021causal}, which explains the difference between a traditional SLMM analysis and an RSR analysis. However, in a Bayesian analysis the posterior distribution is used for inference. Consequently, the equivalence relationships depicted in Figure 1 show that the differences between the data generating model and analysis model are inconsequential for posterior inference under our standard/traditional prior assumptions, and rather, differences in point and interval estimation arise because different quantities are being estimated (i.e., $\bm{\beta}/\bm{\delta}$ are traditionally estimated when using an SLMM, and $\bm{\beta}^{adj}/\bm{\beta}_{RSR}$ can be estimated when using a misspecified RSR).

From Proposition 1, one might be tempted to say that the two step procedure (i.e., first sampling from the misspecified augmented RSR, and then sampling $\bm{\beta}^{adj}$) is simply the process of orthogonalization. However, orthogonalization does not require the use of location invariant priors. It is easily shown that if one reverses the orthoganalization when using a prior that is not location invariant we do not obtain the equivalence relationships in Proposition 1. We provide an example in Appendix B using Gaussian prior assumptions.

The SLMM in (\ref{eq:slmm:bhm2}), that is assumed in Proposition 1, is extremely general and allows for any prior distributional assumptions on $\bm{\Sigma}_{\nu}$ and $\sigma^{2}$. For example, one could consider point mass priors $f(\bm{\Sigma}_{\nu}) = I(\bm{\Sigma}_{\nu} = \bm{\Sigma}_{\nu}^{(0)})$ and $f(\sigma^{2}) = I(\sigma^{2} = \sigma_{0}^{2})$, where $I(\cdot)$ is an indicator function, $\bm{\Sigma}_{\nu}^{(0)}$ is a fixed real-valued positive definite matrix interpreted as the ``true covariance matrix,'' and $\sigma_{0}^{2}>0$ is a fixed value interpreted as the ``true variance.'' The marginal distribution of the data under this degenerate prior distributional assumption becomes the marginal likelihood (i.e., the distribution of the data given $\bm{\Sigma}_{\nu}^{(0)}$ and $\sigma_{0}^{2}$ marginalizing across $\bm{\beta}$ and $\bm{\nu}$), and thus, Proposition 1 shows that $f_{SLMM}(\textbf{y}_{o}\vert \bm{\Sigma}_{\nu}^{(0)},\sigma_{0}^{2}) =  f_{RSR}(\textbf{y}_{o}\vert \bm{\Sigma}_{\nu}^{(0)},\sigma_{0}^{2}) = f_{aRSR}(\textbf{y}_{o}\vert \bm{\Sigma}_{\nu}^{(0)},\sigma_{0}^{2})$ in this degenerate case. Of course, hypothesis testing via the likelihood ratio test would never reject the null under any specification of the null (i.e., the null hypothesis that the likelihood is $f_{RSR}(\textbf{y}_{o}\vert \bm{\Sigma}_{\nu}^{(0)},\sigma_{0}^{2})$ and the null hypothesis that the likelihood is $f_{SLMM}(\textbf{y}_{o}\vert \bm{\Sigma}_{\nu}^{(0)},\sigma_{0}^{2})$ would both produce a likelihood ratio test statistic equal to one). Under our improper prior for $\bm{\beta}$ it is easy to verify, via a completing the squares argument, that this marginal likelihood is a singular normal distribution with location parameter equal to zero and singular precision matrix $(\bm{\Sigma}_{\nu}^{(0)}+\sigma_{0}^{2}\textbf{I})^{-1}
-(\bm{\Sigma}_{\nu}^{(0)}+\sigma_{0}^{2}\textbf{I})^{-1}\textbf{X}(\textbf{X}^{\prime} (\bm{\Sigma}_{\nu}^{(0)}+\sigma_{0}^{2}\textbf{I})^{-1}\textbf{X})^{-1}\textbf{X}^{\prime} (\bm{\Sigma}_{\nu}^{(0)}+\sigma_{0}^{2}\textbf{I})^{-1}$. Thus, standard frequentist strategies assuming a singular covariance matrix, such as the Gauss-Markov theorem to produce the generalized least squares estimator \citep[e.g., see][in the context of the RSR]{zimmerman2022deconfounding}, would be problematic under this (marginal) likelihood assumption.

\subsection{Comparison of the SLMM and the RSR from the Perspective of Impartial Selection Criteria}  The literature supporting Conclusion 1 primarily assumes the SLMM to be the true data generating mechanism, which is sensible considering the construction of the RSR. However, as \citet{hanks2015restricted} suggests, the RSR adds strong assumptions to the SLMM, which ultimately makes it a different model from the SLMM. Thus, when comparing SLMM and the misspecified RSR it is also sensible to follow a more formal model selection perspective, where it is assumed possible that neither the SLMM and/or the RSR represent the distribution of the data. Some people refer to this as the ``M-open'' setting \citep[e.g., see][]{clyde2004model}.

If one is willing to adopt a formal model selection perspective when comparing the RSR and the SLMM, then both the RSR and the SLMM are simply different models, that one might select with a criterion. A fundamental criterion considered in the Bayesian literature is the Bayes factor \citep{gelman2013bayesian}, and in our setting, $BF = \frac{f_{SLMM}(\textbf{y}_{o})}{f_{RSR}(\textbf{y}_{o})}$. It is immediate from Proposition 1 that $BF = 1$ under location invariant priors. This suggests that the SLMM and the RSR are seen as equally appropriate using the Bayes factor for model selection. The Bayes factor is an important criterion that has connections to other impartial selection criteria such as the Bayesian information criterion (BIC) \citep{schwarz1978estimating}. Moreover, from Proposition 1 it is immediate that a Bayesian model average \citep[BMA][]{hoeting1999bayesian} between the SLMM (say $\mathcal{M}_{SLMM}$) and RSR (say $\mathcal{M}_{RSR}$) that places prior probabilities $f(\mathcal{M}_{SLMM}) = f(\mathcal{M}_{RSR})=1/2$ (i.e., considering both models in an impartial manner) produces posterior probability,
\begin{equation*}
	f(\mathcal{M}_{RSR}\vert \textbf{y}_{o}) = \frac{f_{RSR}(\textbf{y}_{o})f(\mathcal{M}_{RSR})}{f_{RSR}(\textbf{y}_{o})f(\mathcal{M}_{RSR})+f_{SLMM}(\textbf{y}_{o})f(\mathcal{M}_{SLMM})} = 1/2 = 	f(\mathcal{M}_{SLMM}\vert \textbf{y}_{o})
\end{equation*}
\noindent
again suggesting that the SLMM and the RSR are equally reasonable using a BMA strategy. Noting that one can write $BF = \frac{f(\mathcal{M}_{RSR}\vert \textbf{y}_{o})f(\mathcal{M}_{SLMM})}{f(\mathcal{M}_{SLMM}\vert \textbf{y}_{o})f(\mathcal{M}_{RSR})}$ we see that $-\mathrm{log}(BF) = 0$ reproduces the difference in the logarithmic score \citep[e.g., see][pg. 233]{wikle2019spatio}. Thus, using the logarithmic score between models we again obtain that we can not distinguish between the RSR and the SLMM. 

Any strictly proper scoring rule \citep{gneiting2007strictly} comparing $f_{SLMM}(\textbf{y}_{m}\vert \textbf{y}_{o})$ to $f_{RSR}(\textbf{y}_{m}\vert \textbf{y}_{o})$ would view both models equally under the assumptions of Proposition 1. That is, a scoring rule $S$, for probability measures $P$ and $Q$, is said to be strictly proper if $S(Q,Q)\ge S(P,Q)$ with equality if and only if $P=Q$. From Proposition 1, we see $P(A) = \int_{A} f_{RSR}(\textbf{y}_{m}\vert \textbf{y}_{o})d\textbf{y}_{m} = \int_{A} f_{SLMM}(\textbf{y}_{m}\vert \textbf{y}_{o})d\textbf{y}_{m} = Q(A)$ for every $A$, and thus, by definition of a strictly proper scoring rule, the SLMM and RSR are equivalent under any strictly proper scoring rule when using location invariant prior distributions. Moreover, it follows immediately from Proposition 1 that the difference between the SLMM's  proper scoring rule with a measure $F$, and the misspecified RSR's  proper scoring rule with a measure $F$ is equal to zero.

The Bayes factor, strictly proper scoring rules, and summaries from the BMA produce metrics that compare distributions, however, several metrics such as the conditional Akaike information criterion \citep{vaida-blanchard}, cross-validation, and leave-one-out cross-validation measure specifically predictive performance \citep{htf}. Proposition 1 shows that the predictive distribution $f_{SLMM}(\textbf{y}_{m}\vert \textbf{y}_{o})=f_{RSR}(\textbf{y}_{m}\vert \textbf{y}_{o})$ under location invariant priors, and as such, an immediate consequence of Proposition 1 is that any impartial metrics that measure predictive summaries (e.g., posterior mean of $\textbf{y}_{m}$) would evaluate both the SLMM and the misspecified RSR equally. Thus, selecting to use the misspecified RSR is equally reasonable as selecting the SLMM according to these criteria.

If one is interested in determining the effect of $\textbf{X}$, I suggest using any one of the aforementioned selection criteria to compare the two versions of the misspecified RSR (based on location invariant priors) that include and exclude a particular covariate. Of course, this is a standard strategy suggested by many and is sometimes referred to as ``best subsets selection'', which is not computationally feasible as $p$ grows. However, several standard strategies exist for large $p$ such as forward selection among others \citep[e.g., see][for a standard reference]{neter1996applied}. Relying on model selection criteria to determine the effect of a covariate avoids having to choose between estimating $\bm{\beta}$ versus $\bm{\delta}$, and allows for both the SLMM and the RSR to be possibly misspecified.  

\section{Robustness Properties for Parameter Estimation}

As described in Section 2.5, to determine the effect of $\textbf{X}$ with a misspecified RSR model, I suggest using selection criteria. However, there are many who would prefer to summarize $\bm{\beta}$ to identify a useful covariate. In this section, we consider estimating parameters in the setting where both the RSR and the SLMM can be misspecified by ignoring unmeasured confounders or through other process model misspecifications. Such misspecifications suggest that the RSR requires adjustment. In particular, we explore the use of the RSR as a Bayesian procedure that can be made robust to unmeasured confounding (after adjustment) for estimating $\bm{\beta}$, $\bm{\delta}$, and $\bm{\theta}$, where $\bm{\theta}$ is a generic $d$-dimensional real-valued parameter vector. Denote an unmeasured confounder with $\textbf{Z}$. By ``robust to unmeasured confounders'' we mean that the following property holds almost surely,
\begin{equation}\label{eq:robust}
	f(\bm{\beta},\bm{\delta},\bm{\theta}\vert \textbf{y},\textbf{X},\textbf{Z}) \approx 	f(\bm{\beta},\bm{\delta},\bm{\theta}\vert \textbf{y},\textbf{X}),
\end{equation}
\noindent
so that posterior inferences on $\bm{\beta}$, $\bm{\delta}$, and $\bm{\theta}$ are unaffected by the knowledge of unmeasured confounders $\textbf{Z}$. More, specifically, our definition of robustness implies that inference on unknown parameters and regression effects are approximately the same when $\textbf{Z}$ is measured (on the left-hand-side of the above equation) and unmeasured (on the right-hand side of the above equation). Robustness of $\bm{\delta}$ to unmeasured confounders was one of the original motivations for the artificial restriction of $\bm{\beta} = \bm{\delta}$ leading to the RSR (Section 3.1). It is important to emphasize that robustness is different from the notion of finding the conditions in which linearly dependent regression effects are identifiable \citep{dupont2022spatial,guan2023spectral,gilbert2025consistency,datta2025consistent}, which is a predominant and important strategy for handling unmeasured confounders, and refers to a property of the likelihood and not the conditional independence in (\ref{eq:robust}).

\subsection{Estimation of Orthogonal Regression Effects in the Presence of Unmeasured Confounders}\label{sec:semi-parametric}

The model selection perspective generally assumes that the probability measure for $\textbf{y}$ is unknown. To perform inference on regression effect, we relax this assumption by assuming certain levels of the hierarchy are unknown and others levels are known. This leads to a more general expression of the additive model (AM) for spatial data, where the model for $\bm{\nu}$ is unknown and potentially influenced by an $n\times c$ matrix $\textbf{Z}$ representing unmeasured confounders. Consider the following hierarchical representation of the semi-parametric AM defined by the product of the following:
\begin{align}
	\nonumber
	f(\textbf{y}\vert \textbf{X}, \bm{\beta},\bm{\nu},\sigma^{2})&= N(\textbf{X}\bm{\beta} + \bm{\nu},\sigma^{2}\textbf{I})\\
	\nonumber
	f(\bm{\beta})  = 1\\
	\label{eq:semi}
	f^{(0)}(\bm{\nu}\vert \textbf{X},\textbf{Z}),
\end{align}
where $f^{(0)}$ denotes the ``true'' pdf used to generate the process $\bm{\nu}$. It is assumed that $\textbf{Z}$ influences both $\textbf{X}$ and $\textbf{y}$ through the models $f^{(0)}(\textbf{X}\vert \textbf{Z})$ and
\begin{equation*}
	f^{(0)}(\textbf{y}\vert \textbf{X},\bm{\beta},\textbf{Z},\sigma^{2}) = \int f(\textbf{y}\vert \textbf{X}, \bm{\beta},\bm{\nu},\sigma^{2}) f^{(0)}(\bm{\nu}\vert \textbf{X},\textbf{Z})d\bm{\nu},
\end{equation*}
\noindent
making $\textbf{Z}$ an unmeasured confounder. For example, consider $c =p= 1$, $f^{(0)}(\bm{\nu}\vert \textbf{X},\textbf{Z})$ can be a point mass on the event $\bm{\nu} = \textbf{Z}$, and $f^{(0)}(\textbf{X}\vert \textbf{Z}) = N(\textbf{Z},2\textbf{I})$ so that both $\textbf{X}$ and $\textbf{y}$ are influenced by $\textbf{Z}$. We say (\ref{eq:semi}) is ``semi-parametric'' because $f^{(0)}$ is assumed to be the ``true'' distribution for $\bm{\nu}\vert \textbf{X},\textbf{Z},\sigma^{2}$. In an effort to avoid confirmation bias, the model for $f^{(0)}(\bm{\nu}\vert \textbf{X},\textbf{Z})$ is allowed to be any specification including $f_{RSR}$, $f_{SLMM}$, or a completely different density that possibly includes unmeasured confounders.

 {In practice, $f^{(0)}$ is unknown and one can consider a potentially misspecifed parametric specification, written hierarchically as,
	\begin{align}
		\nonumber
		f(\textbf{y}\vert \textbf{X}, \bm{\beta},\bm{\nu},\sigma^{2})&= N(\textbf{X}\bm{\beta} + \bm{\nu},\sigma^{2}\textbf{I})\\
		\nonumber
		f(\bm{\beta})  = 1\\
		\label{eq:parametric}
		f(\bm{\nu}\vert \textbf{X},\bm{\theta}),
	\end{align}
	where we specify $\sigma^{2}\in \bm{\theta}$. The model in (\ref{eq:parametric}) may be misspecified as there may not exist a $\bm{\theta}$ such that $	f(\bm{\nu}\vert \textbf{X},\bm{\theta}) = f^{(0)}(\bm{\nu}\vert \textbf{X},\textbf{Z})$.}

Several papers have demonstrated that the marginal predictive distribution for $\bm{\delta}$ in the RSR produces the ordinary least squares (OLS) estimator when $c = 1$ and $\textbf{Z} = \bm{\nu}$ almost surely \citep{rao1967least,Reich,khan2022restricted}. The same property holds true under our more general setting when assuming both the semi-parametric model and the parametric model. \\

\vspace{-20pt}

\noindent
\textbf{Proposition 2:} Let $\bm{\delta} = \bm{\beta} + (\textbf{X}^{\prime}\textbf{X})^{-1}\textbf{X}^{\prime}\bm{\nu}$. {Suppose $\bm{\theta}$ is given a proper prior distribution. Then the predictive distribution for $\bm{\delta}$ derived from the parametric model in (\ref{eq:parametric}):
	\begin{equation}
		\label{eq:OLS}
		f(\bm{\delta}\vert \textbf{y},\textbf{X},\sigma^{2}) = N\left\lbrace(\textbf{X}^{\prime}\textbf{X})^{-1}\textbf{X}^{\prime}\textbf{y},\sigma^{2}(\textbf{X}^{\prime}\textbf{X})^{-1}\right\rbrace.
	\end{equation}
	\noindent
Similarly the predictive distribution for $\bm{\delta}$ derived from the semi-parametric model in (\ref{eq:semi}) is given by:
	\begin{equation}
		\label{eq:OLS2}
		f(\bm{\delta}\vert \textbf{y},\textbf{X},\sigma^{2}) = N\left\lbrace(\textbf{X}^{\prime}\textbf{X})^{-1}\textbf{X}^{\prime}\textbf{y},\sigma^{2}(\textbf{X}^{\prime}\textbf{X})^{-1}\right\rbrace,
	\end{equation}
	\noindent
	with $f(\bm{\delta}\vert \textbf{y},\textbf{X},\sigma^{2})  = f(\bm{\delta}\vert \textbf{y},\textbf{X},\textbf{Z},\sigma^{2})$.}\\

\vspace{-20pt}

\noindent
\textit{Proof:} See the Supplementary Appendix.\\

\vspace{-20pt}

\noindent
A useful bi-product of Proposition 2 is that $\bm{\delta}$ is conditionally independent of {$\textbf{Z}$ given $\textbf{y}$, $\textbf{X}$, and $\sigma^{2}$ when assuming either (\ref{eq:semi}) and (\ref{eq:parametric})}. Of course if one is solely interested in inference on $\bm{\delta}$ then ordinary least squares would be a path of least resistence. However, if you are simultaneously interested in inference on $\bm{\delta}$, spatial prediction, and hyperparameters, then Proposition 1 shows that the misspecified traditional RSR and correctly specified SLMM both give the same posterior inference on these quantities. While of course, one could reasonably estimate $\bm{\beta}$ in this setting, the robustness result suggests that it may be easier to estimate $\bm{\delta}$ as there is no concern of process model misspecification, where incorrect process model specification may influence estimates of $\bm{\beta}$.

{ We emphasize that both the misspecified parametric model and correctly specified semi-parametric model in (\ref{eq:semi}) and (\ref{eq:parametric}) assume the linearly dependent regression coefficients are potentially collinear with the spatial error term $\bm{\nu}$. The main motivation for Proposition 2 is that one can be completely wrong about with $f(\bm{\nu}\vert \textbf{X},\bm{\theta})\ne f^{(0)}(\bm{\nu}\vert \textbf{X},\textbf{Z})$, and the predictive distribution for $\bm{\delta}$ (not $\bm{\beta}$!) is exactly the same as the predictive distribution for $\bm{\delta}$ when assuming the correctly specified model $f^{(0)}$. Thus, the RSR and aRSR, which have the same predictive distribution for $\bm{\beta}_{RSR}$, solves the problem of unmeasured confounding for inference on the orthogonalized regression coefficients (i.e., $\bm{\bm{\delta}}$) when $\sigma^{2}$ is equal to one known value, but does not address the problem of unmeasured confounding for inference on linearly dependent regression coefficients (i.e., $\bm{\beta}$). }

\subsection{Estimation of Linearly Dependent Regression Effects} 

Proposition 2 speaks towards the choice to use summarizes of $\bm{\delta}$ instead of summarizes of $\bm{\beta}$ to investigate the conditional mean of predictive data. In particular, we have two choices to describe the conditional mean of posterior predictive data, namely, $\bm{\delta}$ and $\bm{\beta}$. In frequentist settings, the choice is straightforward as $\bm{\beta}_{0}$ is physically interpretable while $\bm{\delta}_{0}$ is not. However, recall our discussion on data generating mechanisms that imply that both $\bm{\beta}$ and $\bm{\delta}$ do not have any physical interpretations, but rather represent two distinct representations for the conditional mean of posterior predictive data. In this setting, physical interpretation cannot be used to guide our choice to summarize $\bm{\beta}$ or $\bm{\delta}$. The results above show that $\bm{\delta}$ is robust to unmeasured confounders and model misspecification. It is a much more serious mathematical undertaking to show that estimates of $\bm{\beta}$ are identifiable, and existing results require additional assumptions often on the data generating mechanism \citep{dupont2022spatial,gilbert2025consistency,datta2025consistent}. Consequently, when assuming a data generating mechanism with $\bm{\beta}$ random, Proposition 2  provides a semi-parametric justification for choosing to estimate $\bm{\delta}$ instead of $\bm{\beta}$. Of course, if one estimates $\bm{\delta}$ instead of $\bm{\beta}$, they should make it explicitly clear that they estimate orthogonalized regression coefficients and these are different from the linearly dependent regression coefficients.

This perspective is nuanced and ultimately unsatisfying, since the ideal goal is to estimate both $\bm{\beta}$ and $\bm{\delta}$ in the presence of unmeasured confounding. Motivated by this objective, we investigate the two-step strategy proposed by \citep{hanks2015restricted} for estimating linearly dependent regression effects, which I liken to a type of transfer learning.  \citet{hanks2015restricted}'s approach can be motivated by the semiparametric result that $\bm{\delta}$ can be estimated despite process model misspecification. That is, defining a data model with $\bm{\delta}\vert \bm{\beta}^{adj},\bm{\theta},\textbf{y}$ allows us to directly remove process model misspecification that may arise due to unmeasured confounding assuming the semi-parametric additive model. Additionally subsequent prior specifications can make use of expert knowledge and preliminary plug-in estimates. Accordingly, the strategy proceeds by (1) sampling $\bm{\delta}$ from the posterior distribution of a misspecified RSR model and then (2) using the resulting sample of $\bm{\delta}$ to infer $\bm{\beta}^{adj}$. 

Transfer learning can broadly be described as an inferential procedure that takes knowledge gained from learning one task (i.e., Task 1) to improve the performance on a different, but related task (i.e., Task 2). In our context one might consider first estimating $\bm{\delta}$ (Task 1), since its conditional mean is unaffected by unmeasured confounders, and Task 2 could be to use what we've learned about $\bm{\delta}$ to estimate $\bm{\beta}$. This may be an abuse of the term ``transfer learning'' as Task 1 and 2 are traditionally done using different data sources, but we assume the same $\textbf{y}$ and $\textbf{X}$ for both tasks. \citet{hanks2015restricted} applied such a strategy by treating $\bm{\delta}$ as unbiased data for estimating $\bm{\beta}$. In particular, they propose a posterior predictive step, where $\bm{\delta}$ is drawn from $f(\bm{\delta}\vert \textbf{y}_{o})$ and $\bm{\beta}^{adj}_{MoM}$ is drawn from
\begin{equation}\label{eq:tran:post}
	\bm{\beta}^{adj}_{MoM}\vert \bm{\delta},\bm{\Sigma}_{\nu}\sim N\left(\bm{\delta},(\textbf{X}^{\prime}\textbf{X})^{-1}\textbf{X}^{\prime}\bm{\Sigma}_{\nu}\textbf{X}(\textbf{X}^{\prime}\textbf{X})^{-1}\right),
\end{equation}
\noindent
which was motivated by the relation in Section 2.1,
\begin{equation*}
	\bm{\delta}=\bm{\beta}^{adj}_{MoM} + (\textbf{X}^{\prime}\textbf{X})^{-1}\textbf{X}^{\prime}\bm{\nu}.
\end{equation*}
\noindent
We use the subscript ``MoM'' as the mean of $\bm{\beta}^{adj}_{MoM}$ matches the first moment to $\bm{\delta}$. The quantity $\bm{\beta}_{MoM}^{adj}$ is purely a mathematical construct and its distribution assumes conditional independence between $\bm{\beta}^{adj}$ and the data, random effects, and unmeasured confounders given the remaining parameters, i.e.,  $f(\bm{\beta}^{adj}_{MoM}\vert \bm{\delta},\bm{\Sigma}_{\nu},\textbf{X}) = f(\bm{\beta}^{adj}_{MoM}\vert \bm{\delta},\bm{\Sigma}_{\nu}, \bm{\nu},\textbf{X},\textbf{y},\textbf{Z})$. The predictive distribution in (\ref{eq:tran:post}) can be thought of as a ``posterior distribution'' if $\bm{\delta}$ is interpreted as ``data'' in the hierarchical model,
\begin{align*}
	\bm{\delta}\vert \bm{\beta}^{adj}_{MoM},\bm{\Sigma}_{\nu}&\sim N\left(\bm{\beta}^{adj}_{MoM} ,(\textbf{X}^{\prime}\textbf{X})^{-1}\textbf{X}^{\prime}\bm{\Sigma}_{\nu}\textbf{X}(\textbf{X}^{\prime}\textbf{X})^{-1}\right)\\
	f(\bm{\beta}^{adj}_{MoM}) &= 1.
\end{align*}
From a transfer learning perspective it is quite natural to think of $\bm{\delta}$ as a data source for linearly dependent regression effects, as we know that $\bm{\delta}$ is invariant to unmeasured confounders via Proposition 2, and it is also conditionally unbiased for the linearly dependent regression effects via Equation (\ref{eq:changeofvariables}). \citet{hanks2015restricted} considers dropping the assumption that $\bm{\beta} = \bm{\delta}$ and replacing it with $\bm{\beta} =\bm{\beta}^{adj}_{MoM}$ when the RSR is assumed misspecified.

Now, consider a minor adjustment to the strategy in \citet{hanks2015restricted} with
\begin{align*}
	\bm{\delta}\vert \bm{\beta}^{adj}_{trn},\bm{\Sigma}_{\nu}&\sim N\left(\bm{\beta}^{adj}_{trn} ,(\textbf{X}^{\prime}\textbf{X})^{-1}\textbf{X}^{\prime}\bm{\Sigma}_{\nu}\textbf{X}(\textbf{X}^{\prime}\textbf{X})^{-1}\right)\\
	f(\bm{\beta}^{adj}_{trn}) &= N(\bm{\mu}_{\beta},\sigma_{\beta}^{2}\textbf{I}_{p}),
\end{align*}
 \noindent
 which produces the predictive distribution
 \begin{align}
 	\nonumber
 	&f_{trn}(\bm{\beta}^{adj}_{trn}\vert \bm{\delta},\bm{\Sigma}_{\nu},\bm{\mu}_{\beta})\\
 	\nonumber
 	&= N\left\lbrace \left((\textbf{X}^{\prime}\textbf{X})(\textbf{X}^{\prime}\bm{\Sigma}_{\nu}\textbf{X})^{-1}(\textbf{X}^{\prime}\textbf{X})+ \frac{1}{\sigma_{\beta}^{2}}\textbf{I}_{p}\right)^{-1}\left((\textbf{X}^{\prime}\textbf{X})(\textbf{X}^{\prime}\bm{\Sigma}_{\nu}\textbf{X})^{-1}(\textbf{X}^{\prime}\textbf{X})\bm{\delta}+\frac{1}{\sigma_{\beta}^{2}}\bm{\mu}_{\beta}\right),\right.\\
 	\label{eq:tran:post2}
 	&\hspace{50pt}\left.\left((\textbf{X}^{\prime}\textbf{X})(\textbf{X}^{\prime}\bm{\Sigma}_{\nu}\textbf{X})^{-1}(\textbf{X}^{\prime}\textbf{X})+ \frac{1}{\sigma_{\beta}^{2}}\textbf{I}_{p}\right)^{-1}\right\rbrace,
 \end{align}
 where ``trn'' stands for ``transfer.'' We similarly assume conditional independence between $\bm{\beta}^{adj}_{trn}$ and the unmeasured confounders and random effects given the parameters via $f_{trn}(\bm{\beta}^{adj}_{trn}\vert \bm{\delta},\bm{\Sigma}_{\nu},\bm{\mu}_{\beta}(\textbf{y}),\textbf{X}) = f(\bm{\beta}^{adj}_{trn}\vert \bm{\delta},\bm{\Sigma}_{\nu}, \bm{\nu},\bm{\mu}_{\beta},\textbf{X},\textbf{y},\textbf{Z})$.  We allow $\bm{\mu}_{\beta}(\textbf{y})$ to be a function of $\textbf{y}$ (i.e., one can make use of a plug-in estimator). Similar to \citet{hanks2015restricted} we consider dropping the assumption that $\bm{\beta} = \bm{\delta}$ and instead assume $\bm{\beta} =\bm{\beta}^{adj}_{trn}$. From this perspective inference on $\bm{\beta}$ is also a ``restricted'' spatial regression as we consider restricting $\bm{\beta} =\bm{\beta}^{adj}_{trn}$, which is generally not necessarily true and should be seen as an assumption. 

It has been shown that parameters, such as those defining $\bm{\theta}$, can be difficult to learn in several settings \citep{banerjee-etal-2004}. That is, certain values of the hyperparameters defining the prior distribution of $\bm{\theta}$ can imply that $f(\bm{\theta}\vert\textbf{y},\textbf{X},\textbf{Z}) \approx g(\bm{\theta})$ for some density $g$ explicitly leading to no ``Bayesian learning'' (i.e., the posterior distribution is not informed by any data source). This can be explicitly shown for large $n$ for standard models. For example, the spatial range parameter and the noise-to-spatial variance ratio in Bayesian models with Mat\'{e}rn covariances are known to be posterior inconsistent \citep{zhang2004inconsistent}. In Section 3, we consider such a specification. Specifying hyperparameters that leads to no Bayesian learning (unavoidable asymptotically for some standard models), when combined with our transfer learning strategy for $\bm{\beta}_{trn}^{adj}$ in (\ref{eq:tran:post2}) and Proposition 2, produces
\begin{align}
	\nonumber
f(\bm{\beta}_{trn}^{adj},\bm{\delta},\bm{\theta}\vert \textbf{y},\textbf{X},\textbf{Z}) &= f_{trn}(\bm{\beta}_{trn}^{adj}\vert \textbf{y},\textbf{X},\bm{\delta},\bm{\theta})f(\bm{\delta}\vert  \textbf{y},\textbf{X},\textbf{Z},\bm{\theta})f(\bm{\theta}\vert  \textbf{y},\textbf{X},\textbf{Z})\\
\label{eq:empBayes2}
&= f_{trn}(\bm{\beta}_{trn}^{adj}\vert \textbf{y},\textbf{X},\bm{\delta},\bm{\theta})f(\bm{\delta}\vert  \textbf{y},\textbf{X},\sigma^{2})f(\bm{\theta}\vert  \textbf{y},\textbf{X},\textbf{Z})\\
\nonumber
&\approx f_{trn}(\bm{\beta}_{trn}^{adj}\vert \textbf{y},\textbf{X},\bm{\delta},\bm{\theta})f(\bm{\delta}\vert  \textbf{y},\textbf{X},\sigma^{2})g(\bm{\theta})\\
\nonumber
&= f(\bm{\beta}_{trn}^{adj},\bm{\delta},\bm{\theta}\vert \textbf{y},\textbf{X}).
\end{align}
\noindent
While this produces inference that is robust to unmeasured confounders it is not ideal, as we do not learn the values of $\bm{\theta}$, which will in turn effect inference on $\bm{\beta}_{trn}^{adj}$. As such, our suggestion to use an informative prior based on plug-in estimates is particularly useful to incorporate additional information to estimate linearly dependent regression effects. That is, we can leverage existing estimators from the identifiability portion of the literature focused on finding the conditions where one can identify $\bm{\beta}$. In particular, we can make use of strategies that model $\textbf{X}$ directly \citep[e.g., see Spatial+][]{dupont2022spatial}, strategies that focus on the appropriate balance between relative signal-to-noise frequencies \citep{guan2023spectral}, or adopt non-parametric tools such as double-machine-learning \citep{gilbert2021causal}. The hyperparameter $\bm{\mu}_{\beta}$ in our posterior predictive step can be set equal to any of these estimators. Moreover, the identifiability conditions found in the literature \citep{dupont2022spatial,guan2023spectral,gilbert2021causal} hold for our transfer learning strategy as $\sigma_{\beta}^{2}\rightarrow 0$ when using the corresponding estimator as a plug-in the value for $\bm{\mu}_{\beta}$. In our simulations, we consider setting $\bm{\mu}_{\beta}$ equal to Spatial+ and set equal to the double machine learning (DML) shift estimator suggested in \citet{gilbert2021causal}.

\section{A New Posterior Sampling Scheme}  

In this section, we provide new results that show that one specification of the aRSR leads to computational developments in sampling from the posterior distribution. Considering the equivalence relationships described in Figure 1, showing that the aRSR can produce the same posterior inferences as the SLMM, these computational improvements provide a clear practical reason to use the aRSR in contexts where one would use the SLMM.

Note that because $\textbf{P}$ and $\textbf{I}-\textbf{P}$ are idempotent, we know that their eigenvalues are either zero or one. That is, if $\textbf{v}$ is an eigenvector of $\textbf{P}$ then $\textbf{P}\textbf{v} = \ell\textbf{v}$, and $\ell\textbf{v}=\textbf{P}\textbf{v}=\textbf{P}\textbf{P}\textbf{v} = \ell\textbf{P}\textbf{v} = \ell^{2}\textbf{v}$ so that $\ell = \ell^{2}$ implying that $\ell = 0,1$. Moreover, the trace of a matrix is the sum of the eigenvalues, which implies that $\textbf{P}$ has $p$ orthonormal eigenvectors with eigenvalues equal to one, denoted with the $n\times p$ matrix $\textbf{L}_{1}$, and $\textbf{I}-\textbf{P}$ has $n-p$ orthonormal eigenvectors with eigenvalues equal to one, denoted with the $n\times (n-p)$ matrix $\textbf{L}$. It follows that $\textbf{P} = \textbf{L}_{1}\textbf{L}_{1}^{\prime}$, $\textbf{I}-\textbf{P} = \textbf{L}\textbf{L}^{\prime}$, $\textbf{L}_{1}^{\prime}\textbf{L}$ is a $p\times (n-p)$ matrix of zeros, and $(\textbf{L}_{1}, \textbf{L})$ is a $n\times n$ orthonormal matrix. Let $\textbf{L}_{o} = (\bm{0}_{n_{o},n-n_{o}},\textbf{I}_{n_{o}})\textbf{L}$.

Using the spectral decomposition, we can write
\begin{equation}
	\label{eq:sigo}
	\textbf{I}_{n_{o}}+\textbf{L}_{o}\textbf{L}^{\prime}\bm{\Sigma}_{\nu}(\bm{\Lambda})\textbf{L}\textbf{L}_{o}^{\prime} = \textbf{B}\bm{\Lambda}\textbf{B}^{\prime},
\end{equation}
\noindent 
where we consider a decomposition of the matrix $\textbf{I}_{n_{o}}+\textbf{L}_{o}\textbf{L}^{\prime}\bm{\Sigma}_{\nu}(\bm{\Lambda})\textbf{L}\textbf{L}_{o}^{\prime}$ via the $n_{o}\times n_{o}$ orthonormal matrix $\textbf{B}$ is assumed known and pre-specified. Basis function expansions have become a standard tool in spatial and spatio-temporal statistics as they offer a semi-parametric approach to general random process modeling \citep{cressie-wikle-book}. That is, several basis functions are known to be ``complete,'' which refers to representation theorems that show that any generic random process can be represented as an infinite dimensional linear combination of basis functions in $L_{2}$ (e.g., see the Karhunen-Lo\'{e}ve expansion for random process approximations and Mercer's theorem for covariance function approximations). This semi-parametric property allows one to model covariance structures such as the Mat\'{e}rn or the conditional autoregressive model (e.g., see Theorem 1 in \citet{bradleyCAGE}, which provides a basis function representation theorem for areal data). 

 Additionally, let $\bm{\Lambda} = diag(\lambda_{1}+1,\ldots, \lambda_{n_{o}}+1)$ with $\lambda_{j}>0$ for all $j$. The diagonal entries in $\bm{\Lambda}$ shift the $i$-th element 1. This is because the left hand-side is the sum of an identity matrix and a positive-semi-definite matrix, which immediately implies that the eigenvalues are greater than or equal to 1. We consider $\bm{\Sigma}_{\nu}$ to be real matrix-valued function that satisfies (\ref{eq:sigo}). In Section \ref{sec:functional}, we provide a functional form for $\bm{\Sigma}_{\nu}(\bm{\Lambda})$ that satisfies (\ref{eq:sigo}).

This leads to the following special case of the aRSR:
 \begin{align}
 	\nonumber
 	f_{aRSR}(\textbf{y}\vert  \bm{\beta}_{RSR},\bm{\nu},\sigma^{2})&= N(\textbf{X}\bm{\beta}_{RSR} + (\textbf{I}-\textbf{P})\bm{\nu},\sigma^{2}\textbf{I})\\
 	\nonumber
 	f_{RSR}(\bm{\beta}_{RSR})  &= 1\\
 	\nonumber
 	f_{SLMM}(\bm{\nu}\vert \sigma^{2},\bm{\Sigma}_{\nu}(\bm{\Lambda})) &= N(\bm{0}_{n},\sigma^{2}\bm{\Sigma}_{\nu}(\bm{\Lambda}))\\
 	\nonumber
 	f(\sigma^{2}) &= IG(\alpha,\kappa)\\
 	\label{eq:rsr:bhm:aug2}
 	 f(\lambda_{i}^{2}) &= \left(\frac{1}{\lambda_{i}+1}\right)\hspace{2pt}I(0<\lambda_{i}); \hspace{2pt} i = 1,\ldots, n_{o},
 \end{align}
 \noindent
where $IG(\alpha,\kappa)$ is a shorthand for the inverse-gamma distribution with shape $\alpha>p/2$ and rate $\kappa>0$, the indicator function $I(0<\lambda_{i})$ equals 1 when $0<\lambda_{i}$ and is zero otherwise. Re-scaling by $\sigma^{2}$ is a typical assumption and is sometimes referred to as a normal-inverse-gamma model. The signal-to-noise ratio is an important consideration when using the NIG model \citep{zhang2023exact} and for unmeasured confounding \citep{dupont2022spatial}. We will consider a parameterization of $\bm{\Sigma}_{\nu}$ that allows the signal-to-noise to vary freely between zero and infinity. See Section~\ref{sec:functional} for additional details. A power law distribution is assumed for $1+\lambda_{i}$, which is a generally common form in the Bayesian literature for prior distributions on scale parameters \citep{schervish2012theory}. This particular power law distribution will produce a closed form expression of the posterior distribution.

\subsection{Applying the Method of Composition to aRSR}\label{sec:compo}

The posterior distribution of the aRSR in (\ref{eq:rsr:bhm:aug2}) can be decomposed using the method of composition \citep{press2009subjective} as follows,
	\begin{align}
		\nonumber
		&f_{aRSR}(\bm{\beta}^{adj}_{trn},\bm{\beta}_{RSR},\bm{\nu},\sigma^{2},\textbf{y}_{m},\bm{\Lambda}\vert \textbf{y}_{o})\\
		\nonumber
		&= 	f_{trn}(\bm{\beta}^{adj}_{trn}\vert\bm{\beta}_{RSR},\sigma^{2}\bm{\Sigma}_{\nu}(\bm{\Lambda}),\bm{\mu}_{\beta},\textbf{X})f_{aRSR}(\bm{\beta}_{RSR}\vert \textbf{y},\sigma^{2},\textbf{X})\\
		\label{eq:joint:post}
		&\hspace{10pt}\times f_{SLMM}(\bm{\nu}\vert \textbf{y},\sigma^{2},\bm{\Lambda},\textbf{X}) 	f_{aRSR}(\sigma^{2}\vert \textbf{y},\bm{\Lambda},\textbf{X})f_{aRSR}(\textbf{y}_{m}\vert \textbf{y}_{o},\bm{\Lambda},\textbf{X})f_{aRSR}(\bm{\Lambda} \vert\textbf{y}_{o},\textbf{X}),
	\end{align}
	\noindent
	where note that we utilize the conditional independence relationship between $\bm{\beta}_{RSR}$ and $\bm{\nu}$ shown in Proposition 2. Each term in the right-hand-side of (\ref{eq:joint:post}) can be derived in closed form, and several of these densities can be immediately sampled from.\\
	
\noindent
\textbf{Proposition 3:} Assume the hierarchical model in (\ref{eq:rsr:bhm:aug2}). Then the densities  $f_{trn}(\bm{\beta}^{adj}_{trn}\vert\bm{\beta}_{RSR},\sigma^{2}\bm{\Sigma}_{\nu}(\bm{\Lambda}))$, $f_{aRSR}(\bm{\beta}_{RSR}\vert \textbf{y},\sigma^{2})$,
$f_{SLMM}(\bm{\nu}\vert \textbf{y},\sigma^{2},\bm{\Lambda})$,	$f_{aRSR}(\sigma^{2}\vert \textbf{y},\bm{\Lambda}) = IG(\alpha^{*},\kappa^{*})$, and $f_{aRSR}(\textbf{y}_{m}\vert \textbf{y}_{o},\bm{\Lambda})$ are all known in closed form and can be sampled from directly. Additionally,
\begin{align}
	\label{eq:lamb}
	&f(\mathrm{vec}(\bm{\Lambda})\vert \textbf{y}_{o},\textbf{X}) = \frac{\Gamma(\frac{2\alpha-p+n_{o}}{2})}{\Gamma((2\alpha-p)/2)\gamma^{n_{o}/2}(\prod_{i = 1}^{n_{o}}(1/h_{i}^{2}))^{1/2}} \frac{1}{\left(\frac{1}{2\alpha-p}\sum_{k = 1}^{n_{o}}\frac{h_{k}^{2}}{\lambda_{k}+1}+1\right)^{\frac{n_{o}+2\alpha-p}{2}}} \prod_{i = 1}^{n_{o}}\left(\frac{1}{\lambda_{i}+1}\right)^{3/2},
\end{align}
\noindent
where the function ``vec$(\bm{\Lambda})$'' produces the vector along the main diagonal of $\bm{\Lambda}$, $\textbf{h} = (h_{1},\ldots,h_{n_{o}})^{\prime} = \left(\frac{2\alpha - p}{2\kappa}\right)^{1/2}\textbf{B}^{\prime}\textbf{y}_{o}$, $\alpha^{*} = (n-p)/2+\alpha$, and $\kappa^{*} = \textbf{y}^{\prime}\left((\textbf{I}-\textbf{P})\bm{\Sigma}_{\nu}(\textbf{I}-\textbf{P})+\textbf{I}\right)^{-1}\textbf{y}/2 + \kappa$. \\

\noindent
\textit{Proof:} For a more formal statement and proof see the Supplementary Appendix.\\

\noindent
Sampling from $f(vec(\bm{\Lambda})\vert \textbf{y}_{o})$ is not immediate from Proposition 3. This can be achieved through a change-of-variables to a truncated multivariate $t$ distribution, which we formally state in Proposition 4. \\

\noindent
\textbf{Proposition 4:} Suppose $vec(\bm{\Lambda})$ is distributed according to $f(vec(\bm{\Lambda})\vert \textbf{y}_{o})$ in (\ref{eq:joint:post}). Let $\textbf{g} = (g_{1},\ldots, g_{n_{o}})^{\prime}$ be distributed as a truncated multivariate $t$-distribution with support $[-1,1]$, mean zero, covariance matrix $\textbf{H}$, and degrees of freedom $2\alpha-p$, where $\textbf{H} = \mathrm{diag}(\frac{1}{h_{1}^{2}},\ldots, \frac{1}{h_{n_{o}}^{2}})$. Then $\lambda_{i}$ is equal in distribution to $1/g_{i}^{2} - 1$ for $i = 1,\ldots, n_{o}$.\\

\noindent
\textit{Proof:} See the Supplementary Appendix.\\

\noindent
To simulate directly from the truncated multivariate-$t$ distribution we use the efficient exponential tilting algorithm from \citet{botev2015efficient}. We provide an outline of our entire sampling scheme in Algorithm 1.

When adopting the aRSR in (\ref{eq:rsr:bhm:aug2}) we see that ``small'' values of $\alpha$ and for large $n$ we obtain marginal posterior inference that is robust to unmeasured confounders. That is posterior samples of covariance parameters may be interpreted as draws from a distribution $g(\bm{\theta})$ that does not depend on any data source. In particular, if we assume the true conditional density for $\bm{\nu}$ denoted $f^{(0)}$ in (\ref{eq:semi}) implies $\textbf{y}^{\prime}\left((\textbf{I}-\textbf{P})\bm{\Sigma}_{\nu}(\textbf{I}-\textbf{P})+\textbf{I}\right)^{-1}\textbf{y}/n$ converges almost surely to a finite value, say $\sigma_{0}^{2}>0$ \citep[e.g., see][for relevant related results]{xi2020convergence}, then for large $n$, ${\kappa}^{*} \approx n \sigma_{0}^{2}/2 + \kappa$. From Propositions 3 and 4, it follows that for large $n$, $f(\sigma^{2}\vert \textbf{y},\textbf{Z},\textbf{X}) \approx IG(n/2,n\sigma_{0}^{2} + \kappa)$, and for small $\alpha$, $\lambda_{i}$ is roughly distributed independently according to $g(\lambda_{i}\vert \textbf{y}_{o})=\frac{1}{2(1+\lambda_{i})^{3/2}}I(\lambda_{i}>0)$. Both $IG(n/2,n\sigma_{0}^{2} + \kappa)$ and $g(\lambda_{i}\vert \textbf{y}_{o})=\frac{1}{2(1+\lambda_{i})^{3/2}}I(\lambda_{i}>0)$ do not make use of the realized value of $(\textbf{y},\textbf{X},\textbf{Z})$. Following Equation (\ref{eq:empBayes2}), we see that such a specification leads to posterior inference that are robust to the realized unmeasured confounders.

The normal-inverse-gamma structure, and the additional specifications in Section 4.2, are all non-standard. From the perspective of spatial confounding this suggests that additional consideration of the signal-to-noise ratio (SNR) is needed as many have shown that this is a crucial consideration \citep{guan2023spectral}. Let $\bm{\Lambda}_{-I} = diag(\lambda_{1},\ldots, \lambda_{n_{o}})$ so that $\bm{\Lambda} = \bm{\Lambda}_{-I} + \textbf{I}_{n_{o}}$. The aRSR assumed in Equation (\ref{eq:rsr:bhm:aug2}) implies a signal to noise ratio (SNR) for the observed data $\textbf{y}_{o}$ as,
\begin{equation*}
	SNR = \frac{||cov\{\textbf{L}_{o}\textbf{L}^{\prime}\bm{\nu}\vert \bm{\beta}_{RSR},\textbf{X},\bm{\Lambda},\sigma^{2}\}||_{F}^{2}}{||\sigma^{2}\textbf{I}_{n_{o}}||_{F}^{2}}=\frac{||\textbf{B}\bm{\Lambda}_{-I}\textbf{B}^{\prime}||_{F}^{2}}{||\sigma^{2}\textbf{I}||_{F}^{2}} = \frac{\sum_{i = 1}^{n_{o}}\lambda_{i}^{2}}{n_{o}}.
\end{equation*}
\noindent
Thus, $\lambda_{min}^{2}\le SNR \le \lambda_{max}^{2}$, where $\lambda_{min} = \mathrm{arg\hspace{2pt}min}(\lambda_{1},\ldots, \lambda_{n})$ and $\lambda_{max} = \mathrm{arg\hspace{2pt}max}(\lambda_{1},\ldots, \lambda_{n})$. From Proposition 4, each posterior replicate of $\lambda_{i}\in (0,\infty)$, which suggests that posterior replicates of $\lambda_{min},\lambda_{max}\in (0,\infty)$ and hence the posterior distribution allows $SNR \in (0, \infty)$. Consequently, our non-standard parametric structure does not impose any constraints on the signal to noise ratio.

\begin{algorithm}[htp]
	\caption{Implementation of RSR assuming $\bm{\beta} = \bm{\delta}$, aRSR assuming $\bm{\beta} = \bm{\beta}^{adj}_{MoM}$, aRSR assuming $\bm{\beta} = \bm{\beta}^{adj}_{trn}$, and the SLMM in (\ref{eq:sigo}).}
	\label{aRSR}
	\vspace{10pt}
\begin{enumerate}
	\item Set $w = 1$
	\item Sample $\bm{\Lambda}^{[w]}$ from $f_{aRSR}(\mathrm{vec}(\bm{\Lambda})\vert \textbf{y}_{o})$.
	\item Sample $\textbf{y}_{m}^{[w]}$ from $f_{aRSR}(\textbf{y}_{m}\vert \textbf{y}_{o},\bm{\Lambda}^{[w]})$. Compute $\textbf{y}^{[w]} = (\textbf{y}_{o}^{\prime},\textbf{y}_{m}^{[w]\prime})$.
	\item Sample $\sigma^{2[w]}$ from $f_{aRSR}(\sigma^{2}\vert \textbf{y}^{[w]},\bm{\Lambda}^{[w]})$.
	\item Sample $\bm{\nu}^{[w]}$ from $f_{aRSR}(\bm{\nu}\vert \textbf{y}^{[w]},\sigma^{2[w]},\bm{\Lambda}^{[w]})$ and $\bm{\delta}^{[w]}$ from $f_{aRSR}(\bm{\beta}_{RSR}\vert \textbf{y}^{[w]},\sigma^{2[w]})$ in parallel.
	\item Repeat Steps 1 $\--$ 5 $W$ times.
	\item To implement the SLMM via Section \ref{sec:Conclusion 3}, for each $w$, set $\bm{\beta}^{adj[w]} = \bm{\delta}^{[w]} - (\textbf{X}^{\prime}\textbf{X})^{-1}\textbf{X}^{\prime}\bm{\nu}^{[w]}$.
	\item For each $w$ sample $\bm{\beta}_{MoM}^{adj[w]}$ from $N(\bm{\delta}^{[w]},\sigma^{2[w]}(\textbf{X}^{\prime}\textbf{X})^{-1}\textbf{X}^{\prime}\bm{\Sigma}_{\nu}(\bm{\Lambda}^{[w]})\textbf{X}(\textbf{X}^{\prime}\textbf{X})^{-1})$.
	\item For each $w$ sample $\bm{\beta}_{trn}^{adj[w]}$ from 
	$N\left\lbrace \left(\frac{1}{\sigma^{2[w]}}(\textbf{X}^{\prime}\textbf{X})(\textbf{X}^{\prime}\bm{\Sigma}_{\nu}(\bm{\Lambda}^{[w]})\textbf{X})^{-1}(\textbf{X}^{\prime}\textbf{X})+ \frac{1}{\sigma_{\beta}^{2}}\textbf{I}_{p}\right)^{-1}\right.$\\
	$\hspace{100pt}\left(\frac{1}{\sigma^{2[w]}}(\textbf{X}^{\prime}\textbf{X})(\textbf{X}^{\prime}\bm{\Sigma}_{\nu}(\bm{\Lambda}^{[w]})\textbf{X})^{-1}(\textbf{X}^{\prime}\textbf{X})\bm{\delta}^{[w]} + \frac{1}{\sigma_{\beta}^{2}}\bm{\mu}_{\beta}\right)$,\\
	$\hspace{100pt}\left.\left(\frac{1}{\sigma^{2[w]}}(\textbf{X}^{\prime}\textbf{X})(\textbf{X}^{\prime}\bm{\Sigma}_{\nu}(\bm{\Lambda}^{[w]})\textbf{X})^{-1}(\textbf{X}^{\prime}\textbf{X})+ \frac{1}{\sigma_{\beta}^{2}}\textbf{I}_{p}\right)^{-1}\right\rbrace$.
\end{enumerate}
\end{algorithm}

\subsection{Specification of $\bm{\Sigma}_{\nu}(\bm{\Lambda})$}\label{sec:functional}

Consider the following specification:
\begin{equation}\label{eq:param}
	\bm{\Sigma}_{\nu} = \bm{\Phi}\textbf{M}\bm{\Phi}^{\prime} + \epsilon\textbf{I},
\end{equation}
where the $n\times n_{o}$ matrix $\bm{\Phi}$ is a complete class of spatial basis functions (e.g., Fourier basis, splines, wavelets, etc), $\textbf{M}$ is a positive definite matrix, and $\epsilon>0$ is positive real scalar. By complete we mean that as $n_{o}$ goes to infinity linear combinations of $\bm{\Phi}$ can approximate an arbitrary function in $L_{2}$ \citep[][pg. 102]{cressie-wikle-book}. This parametric form is full rank and $\textbf{P}\bm{\Phi}$ is not necessarily equal to a zero matrix.

Let the polar decomposition of the $n_{o}\times n_{o}$ matrix $\textbf{L}_{o}\textbf{L}^{\prime}\bm{\Phi}$ be $\textbf{B}\textbf{R}$, where $\textbf{B}$ is orthonormal and $\textbf{R}$ is positive definite. Note that the matrix $\textbf{B}$ is completely defined by our choice of $\bm{\Phi}$ and $\textbf{L}_{o}\textbf{L}^{\prime}$ computed from the given matrix $\textbf{X}$. Also, let
\begin{align}\nonumber
	\textbf{M} &= \textbf{R}^{-1}(\bm{\Lambda}_{-I} - \epsilon \textbf{B}^{\prime}\textbf{L}_{o}\textbf{L}_{o}^{\prime}\textbf{B})\textbf{R}^{-1\prime}\\
	\nonumber
	&=  \textbf{R}^{-1}\bm{\Lambda}_{-I}^{1/2}(\textbf{I}_{n_{o}} - \epsilon \bm{\Lambda}_{-I}^{-1/2\prime}\textbf{B}^{\prime}\textbf{L}_{o}\textbf{L}_{o}^{\prime}\textbf{B}\bm{\Lambda}_{-I}^{-1/2\prime})\bm{\Lambda}_{-I}^{1/2\prime}\textbf{R}^{-1\prime}\\
	\nonumber
	&=  \textbf{R}^{-1}\bm{\Lambda}_{-I}^{1/2}(\textbf{I}_{n_{o}} - \epsilon \bm{\Phi}_{M}\bm{\Lambda}_{M}\bm{\Phi}_{M}^{\prime})\bm{\Lambda}_{-I}^{1/2\prime}\textbf{R}^{-1\prime}\\
	\label{eq:M:signu}
	&=\textbf{R}^{-1}\bm{\Lambda}_{-I}^{1/2}\bm{\Phi}_{M}(\textbf{I}_{n_{o}} - \epsilon \bm{\Lambda}_{M})\bm{\Phi}_{M}^{\prime}\bm{\Lambda}_{-I}^{1/2\prime}\textbf{R}^{-1\prime},
\end{align}
where the spectral decomposition of the positive-semi-definite matrix $\bm{\Lambda}^{-1/2\prime}\textbf{B}^{\prime}\textbf{L}_{o}\textbf{L}_{o}^{\prime}\textbf{B}\bm{\Lambda}^{-1/2\prime}=\bm{\Phi}_{M}\bm{\Lambda}_{M}\bm{\Phi}_{M}^{\prime}$, the $n_{o}\times n_{o}$ matrix $\bm{\Phi}_{M}$ is orthonormal, $\bm{\Lambda}_{M} = diag(\lambda_{1,M},\ldots, \lambda_{n_{o},M})$, and $\lambda_{1,M}\ge \ldots \ge \lambda_{n_{o},M}\ge 0$. Let $\epsilon\equiv 1/\lambda_{1,M}$ so that $(\textbf{I}_{n_{o}} - \epsilon \bm{\Lambda}_{M})$ has strictly positive diagonal entries, which implies from (\ref{eq:M:signu}) that $\textbf{M}$ is positive definite. Substituting $\textbf{M}$ in (\ref{eq:param}), and (\ref{eq:param}) into (\ref{eq:sigo}) produces
	\begin{align*}
		\textbf{I}_{n_{o}}+\textbf{L}_{o}\textbf{L}^{\prime}(\bm{\Phi}\textbf{M}\bm{\Phi}^{\prime} + \epsilon\textbf{I})\textbf{L}\textbf{L}_{o}^{\prime} &= 		\textbf{I}_{n_{o}}+\epsilon\textbf{L}_{o}\textbf{L}_{o}^{\prime}+\textbf{B}\textbf{R}\textbf{M}\textbf{R}^{\prime}\textbf{B}^{\prime}\\
		&= 		\textbf{I}_{n_{o}}+\epsilon\textbf{L}_{o}\textbf{L}_{o}^{\prime}+\textbf{B}\textbf{R}\textbf{R}^{-1}(\bm{\Lambda}_{-I} - \epsilon \textbf{B}^{\prime}\textbf{L}_{o}\textbf{L}_{o}^{\prime}\textbf{B})\textbf{R}^{-1\prime}\textbf{R}^{\prime}\textbf{B}^{\prime}\\
		& =\textbf{I}_{n_{o}}+\epsilon\textbf{L}_{o}\textbf{L}_{o}^{\prime}+\textbf{B}\bm{\Lambda}_{-I}\textbf{B}^{\prime} - \epsilon\textbf{L}_{o}\textbf{L}_{o}^{\prime}\\
		&= \textbf{B}\bm{\Lambda}\textbf{B}^{\prime}.
	\end{align*}
	\noindent
	Thus, we see that $\bm{\Sigma}_{\nu}$ in (\ref{eq:param}) with $\textbf{M}$ defined in (\ref{eq:M:signu}) and $\epsilon = 1/\lambda_{1,M}$ satisfies our restriction stated in (\ref{eq:sigo}). Moreover, $\bm{\Phi}\textbf{M}^{1/2}$ defines a complete bases, allowing semi-parametric inference \citep{daw2022overview}.

		\section{Illustrations}
		
		\subsection{Simulation Study}
		\label{sec:verify}

In this section, we illustrate the use of our new adjusted RSR model using a modified version of the simulation model used in \citet{paciorek2010importance}. In particular, we consider the case where the data is simulated according to
\begin{equation*}
	Y(\textbf{s}_{i}) = \beta_{0} + \beta_{x} x(\textbf{s}_{i}) + \beta_{z} z(\textbf{s}_{i}) + \epsilon(\textbf{s}_{i}),
\end{equation*}
\noindent
where $\textbf{s}_{i}$ is the $i$-th location in $n$ values simulated within the unit square and $\{\epsilon(\textbf{s}_{i})\}$ are iid standard normal random variables. We further assume $\textbf{X} = (x(\textbf{s}_{1}),\ldots, x(\textbf{s}_{n}))^{\prime} = \textbf{X}_{c} + \textbf{X}_{u}$, where $\mathrm{cov}(\textbf{Z}) = \textbf{R}(\theta_{c})$, $\mathrm{cov}(\textbf{X}_{c}) = \textbf{R}(\theta_{c})$, $\mathrm{cov}(\textbf{X}_{u}) = \textbf{R}(\theta_{u})$, $\mathrm{cov}(\textbf{X}_{c},\textbf{Z}) = \rho\textbf{R}(\theta_{c})$, and $\textbf{R}(\theta)$ is a $n\times n$ correlation matrix with elements representing the Mat\'{e}rn correlation function with range parameter $\theta$ and smoothness parameter $\nu = 2$. This is one of many simulation specifications that are meant to represent the ``base'' setting in \citet{paciorek2010importance}. We deviate from the general setup from \citet{paciorek2010importance}, where we assume that $\beta_{x}$ and $\beta_{0}$ are independently simulated from a normal distribution with mean zero and variance 3. This is meant to be more reflective of our setting where our data generating mechanism implies that the regression coefficients are random. We randomly select 10$\%$ of the locations to be missing, and in this section present the results for $n = 100$, choose the signal-to-noise ratio to be 2, and $\rho = 0.9$.

We implement the aRSR model in Section 4 with $\bm{\Phi}$ specified to be B-splines, $\alpha = p/2 + 0.5$, $\kappa = 0.5$, and when sampling $\bm{\beta}^{adj}$ we set $\sigma_{\beta}^{2} = 0.1$ and $\bm{\mu}_{\beta}$ to the estimates, Spatial+, DML, or OLS. To implement the aRSR we do not require MCMC as one can sample directly from the posterior distribution, and 200 independent replicates are sampled from the posterior distribution. Consider the root mean squared error for $\bm{\delta}$ and $\bm{\beta}$, and the mean squared prediction error to evaluate orthogonal and linearly dependent regression effects. That is,
$\mathrm{RMSE}_{\beta}(\widehat{\bm{\beta}}) = \left\lbrace \left(\bm{\beta} - \widehat{\bm{\beta}}\right)^{\prime}\left(\bm{\beta}- \widehat{\bm{\beta}}\right)/2\right\rbrace^{1/2}$, coverage over 50 independent replicate simulations, and the root mean square prediction error (RMPSE) between the latent process $\textbf{X}\bm{\beta} + \bm{\nu}$ and the corresponding posterior mean. Interval estimates for Spatial+ and DML follow the bootstrap strategies described in \citet{dupont2022spatial} and \citet{gilbert2021causal}, and Bayesian strategies use 95$\%$ pointwise credible intervals. The RMSPE is averaged over missing locations. We consider $\widehat{\bm{\beta}}$ set equal to SLMM's posterior mean of linearly dependent regression effects, RSR's posterior mean, Spatial+, DML (based on thin plate splines), and the Spatial+, DML, and OLS specifications of our new transfer learning approach $E(\bm{\beta}^{adj}_{trn}\vert \textbf{y}_{o})$.

\begin{figure}[htbp]
	\centering
	\includegraphics[width=0.8\textwidth]{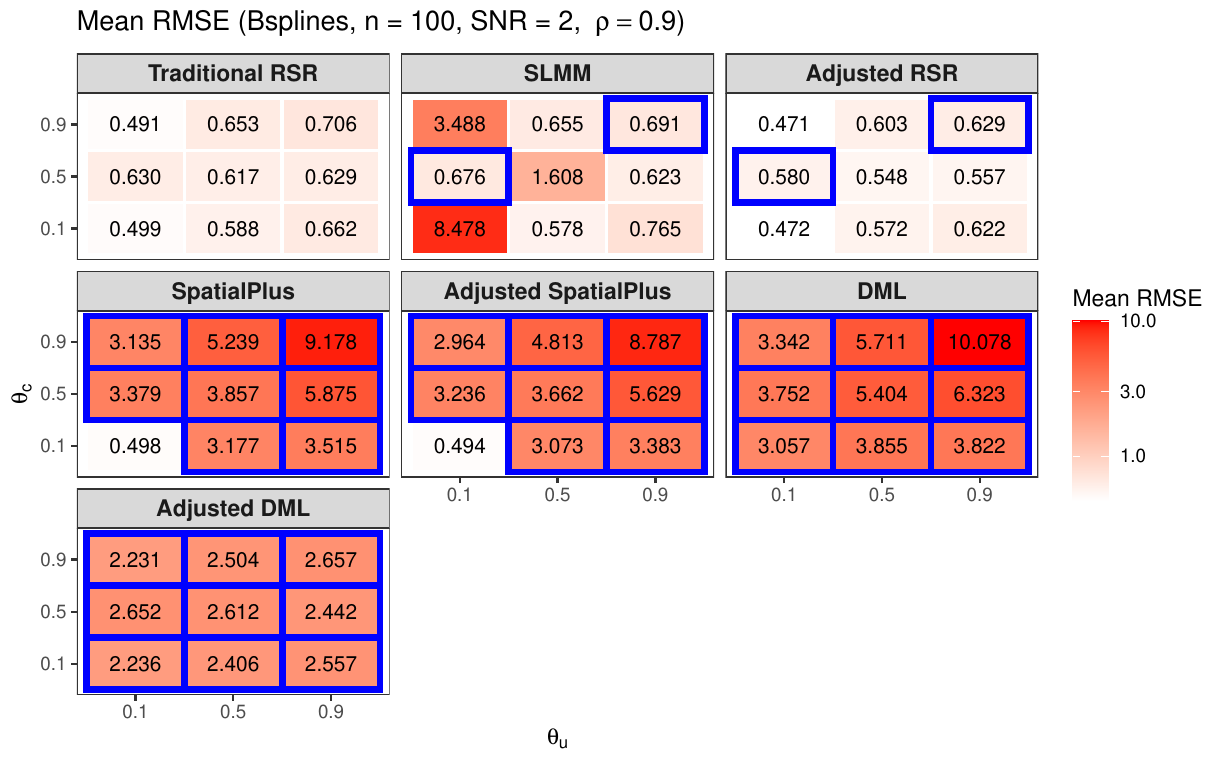}
	\caption{The average $\mathrm{RMSE}_{\beta}$ over 50 simulated replicate datasets at SNR = 2 and $\rho = 0.9$. The $y$-axis considers different values of $\theta_{c}$, and the $x$-axis considers different values of $\theta_{u}$. All other simulation specifications are listed in Section~\ref{sec:verify}. The panel title ``Adjusted RSR'' corresponds to estimates derived from the posterior mean of $\bm{\beta}_{trn}^{adj}$ setting $\bm{\mu}_{\beta}$ equal to the OLS estimator. The panel title ``Adjusted SpatialPlus'' corresponds to estimates derived from the posterior mean of $\bm{\beta}_{trn}^{adj}$ setting $\bm{\mu}_{\beta}$ equal to the Spatial+ estimator. The panel title ``Adjusted DML'' corresponds to estimates derived from the posterior mean of $\bm{\beta}_{trn}^{adj}$ setting $\bm{\mu}_{\beta}$ equal to the DML shift estimator.  Blue cells indicate that a pairwise $t$ test indicates a significant difference between the respective SLMM/Adjusted RSR, SpatialPlus/Adjusted SpatialPlus and DML/Adjusted DML panels.}
	\label{fig:rmse}
\end{figure}

\begin{figure}[htbp]
	\centering
	\includegraphics[width=0.8\textwidth]{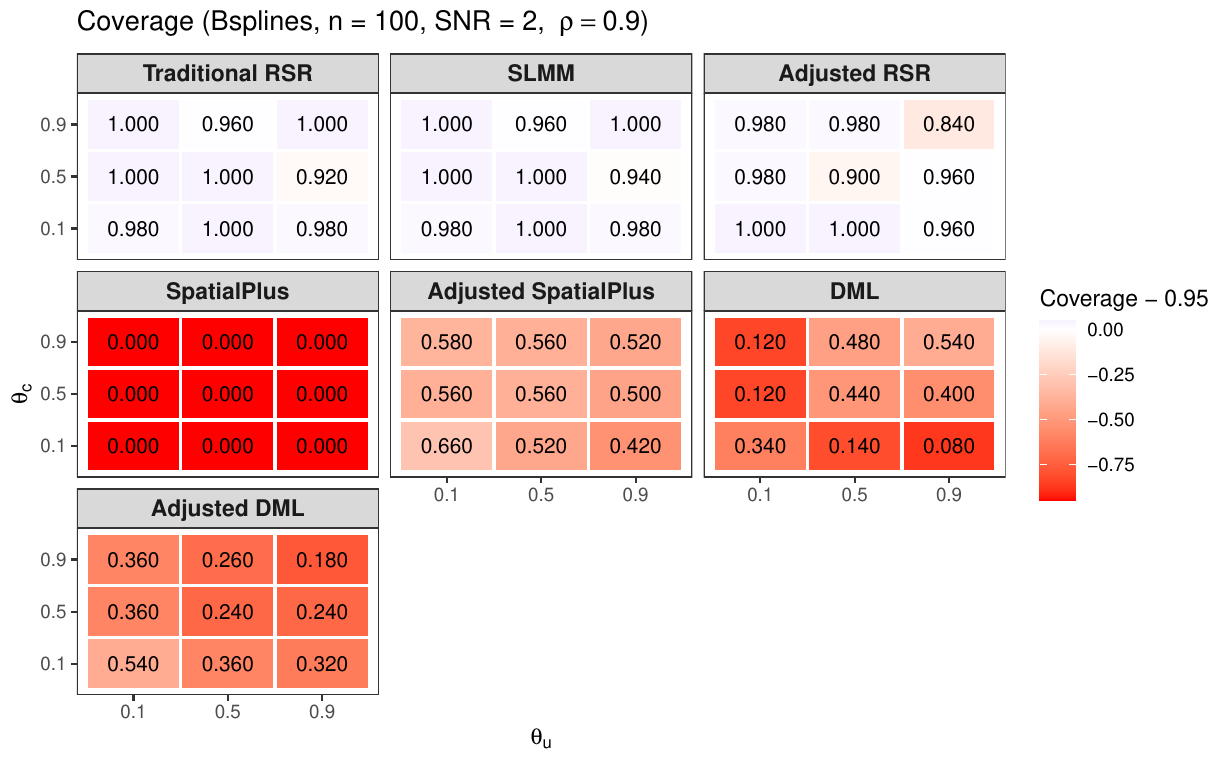}
	\caption{The coverage of $\bm{\beta}$ over 50 simulated replicate datasets at SNR = 2 and $\rho = 0.9$. The $y$-axis considers different values of $\theta_{c}$, and the $x$-axis considers different values of $\theta_{u}$. All other simulation specifications are listed in Section~\ref{sec:verify}. The panel title ``Adjusted RSR'' corresponds to estimates derived from the posterior mean of $\bm{\beta}_{trn}^{adj}$ setting $\bm{\mu}_{\beta}$ equal to the OLS estimator. The panel title ``Adjusted SpatialPlus'' corresponds to estimates derived from the posterior mean of $\bm{\beta}_{trn}^{adj}$ setting $\bm{\mu}_{\beta}$ equal to the Spatial+ estimator. The panel title ``Adjusted DML'' corresponds to estimates derived from the posterior mean of $\bm{\beta}_{trn}^{adj}$ setting $\bm{\mu}_{\beta}$ equal to the DML shift estimator. }
	\label{fig:coverage}
\end{figure}

\begin{figure}[htbp]
	\centering
	\includegraphics[width=0.8\textwidth]{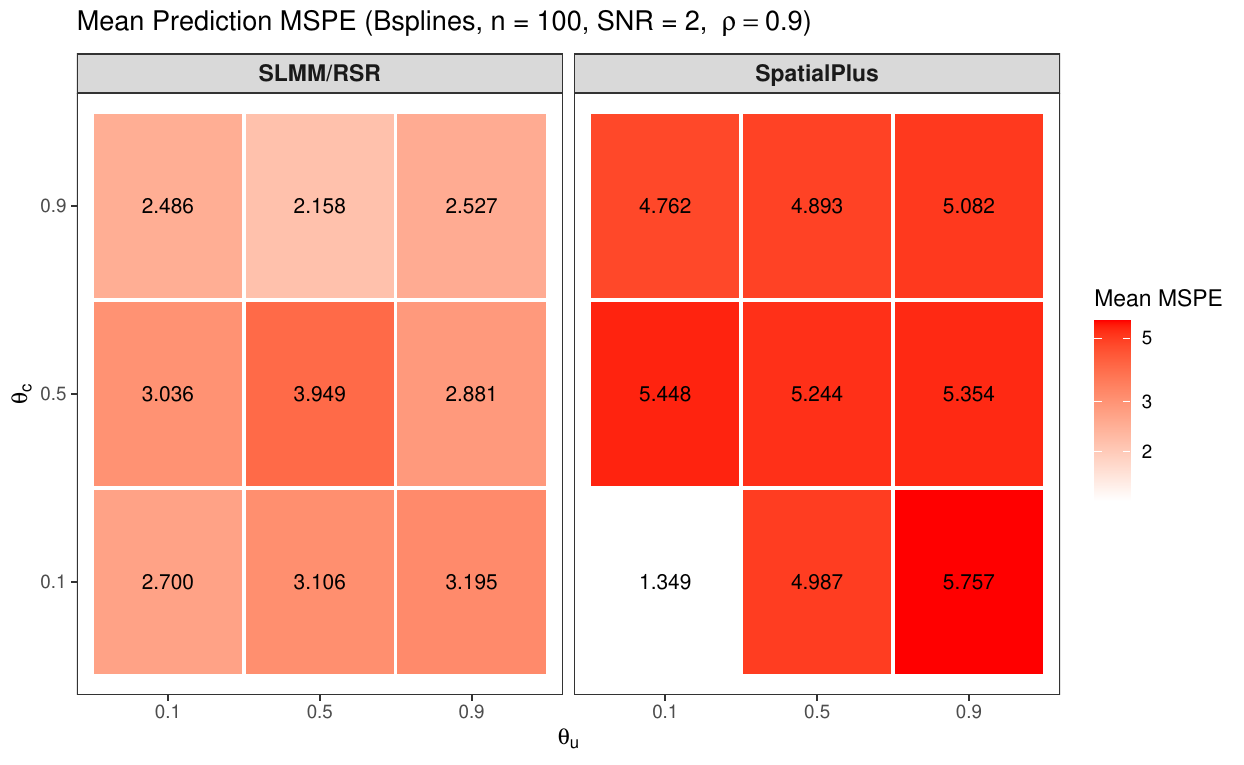}
	\caption{The average $\mathrm{RMSPE}$ over 50 simulated replicate datasets at SNR = 2 and $\rho = 0.9$. The $y$-axis considers different values of $\theta_{c}$, and the $x$-axis considers different values of $\theta_{u}$. All other simulation specifications are listed in Section~\ref{sec:verify}. The panel title ``Adjusted RSR'' corresponds to estimates derived from the posterior mean of $\bm{\beta}_{trn}^{adj}$ setting $\bm{\mu}_{\beta}$ equal to the OLS estimator. The panel title ``Adjusted SpatialPlus'' corresponds to estimates derived from the posterior mean of $\bm{\beta}_{trn}^{adj}$ setting $\bm{\mu}_{\beta}$ equal to the Spatial+ estimator. The panel title ``Adjusted DML'' corresponds to estimates derived from the posterior mean of $\bm{\beta}_{trn}^{adj}$ setting $\bm{\mu}_{\beta}$ equal to the DML shift estimator.  All respective cells produce a pairwise $t$-test that is significant.}
	\label{fig:rmspe}
\end{figure}

		Figures 1$\--$3 provides the summaries of RMSE, coverage, and RMSPE, respectively. In each figure we vary  $\theta_{c}$ and $\theta_{u}$. In general, our proposed adjusted RSR with $\bm{\mu}_{\beta}$ set equal to the ordinary least squares estimators outperforms all methods in terms of MSE, and surprisingly, this is followed by the traditional use of the RSR. The primary deviation in our simulation setup is that we generate $\bm{\beta}$ from a distribution (i.e., a normal mean zero and variance 3), which may explain why estimates of $\bm{\delta}$ outperform Spatial+ and DML as an estimate of $\bm{\beta}$ in this setting. This is an indication that our data generating mechanism that assumes $\bm{\beta}$ is random is important. This pattern is present except for when $\theta_{c} = \theta_{u} = 0.1$, where Spatial+ outperforms all methods except for the adjusted RSR with $\bm{\mu}_{\beta}$ set equal to the estimate from Spatial+. In fact, we generally see that the adjusted RSR that sets $\bm{\mu}_{\beta}$ equal to an estimator outperforms that estimator in terms of the RMSE and coverage. Spatial+, DML, and the adjusted RSR centered on Spatial+ and DML all severely undercover in this simulation design, with clear improvements when using the corresponding adjusted RSR. The misspecified RSR, SLMM, and adjusted RSR centered on OLS all over-cover with the adjusted RSR centered on OLS closer to the nominal 0.95. We note that credible intervals have no guarantee of frequentist coverage. In terms of prediction, the posterior mean from the SLMM/RSR has smaller MSPE than Spatial+ over missing locations except for the case $\theta_{c} = \theta_{u} = 0.1$. 
		
We repeated this study for $n = 200$, the use of thin-plate-splines, a signal-to-noise ratio of 10, and $\rho = 0.1, 0.5,0.9$. We provide 3 additional figures in Supplemental Appendix C (remaining 18 figures not shown). The SLMM produced unstable estimates of $\bm{\beta}$ leading to severe overcoverage, although RMSE values tended not to be significantly different (via paired $t$ test) from the adjusted RSR centered at the OLS. In general, we see similar patterns across each combination of factors, as in Figures 1$\--$3, with the additional result that as $\rho$ decreases all strategies except the SLMM (which over-covered) tend to undercover with the traditional RSR followed by the adjusted RSR centered at the OLS with more preferable coverage rates.

	\begin{table}[t!]
	\centering
		\begin{tabular}{lcc}
			Method & Lower Bound & Upper Bound\\ \hline
			SLMM & -80.96 & 63.05\\
			aRSR ($\bm{\delta}$)& 1.93 & 16.31\\
			aRSR ($\bm{\beta}^{adj}_{trn}$)&6.72 & 13.30
		\end{tabular}
		\caption{The lower and upper bound on the 95$\%$ credible interval for the linearly dependent regression coefficient and orthogonal regression coefficient associated with PM2.5 }\label{tab:cred}
		\end{table}
\subsection{Application to PM2.5 and COVID-19 Mortality}

To illustrate the benefits of aRSR consider data from \citet{wu2020exposure}. Specifically, we consider U.S. counties where a ``large'' mortaility was recorded up to April 22, 2020. Large count-valued responses are arguably normally distributed (e.g., via the central limit theorem), and  by ``large'' we mean counties with 5 or more observed moralities. As an illustration, we only include an intercept and average fine particulate matter (PM2.5) to define the columns of $\textbf{X}$ in our analysis. There are certainly many confounders that would explain COVID-19 mortality including pre-existing health conditions, age, and the use of vaccination \citep{dangi2020review}. Consequently, we would expect $\bm{\beta}$ to potentially be affected by the presence of unmeasured confounders, but from Proposition 2 we expect $\bm{\delta}$ to be unaffected. The 690 $\times$ 690 matrix $\textbf{B}$ is specified to be the eigenvectors of the precision matrix in an intrinsic conditional autoregressive model \citep{besag-74}, and we set $\bm{\mu}_{\beta}$ to the ordinary least squares estimator and $\sigma_{\beta}^{2} = 3$. We plot the predicted versus observed log (for visualization purposes) mortality in Figure~\ref{fig:pred:aRSR}, and see that our predictions track the data fairly well with more variability in small-valued counts. Additionally, in Table~\ref{tab:cred} we provide credible intervals for the linearly dependent regression effect and the orthogonal regression effect associated with PM2.5. The SLMM suggests that PM2.5 was not significant (i.e., zero is in the credible interval), however, as stated above this counter-intuitive conclusion may be due to the presence of known unmeasured confounders. However, both $\bm{\delta}$ and $\bm{\beta}^{adj}_{trn}$ suggest that PM2.5 is significant. From Proposition 2, we see that $\bm{\delta}$ is invariant to unmeasured confounders, and when using it as a data source to produce inference via $\bm{\beta}^{adj}_{trn}$, we see that we are still able to determine that PM2.5 is significant despite the fact that there are unmeasured confounders. From our simulation studies, we see that it is possible that the RSR and the adjusted RSR credible intervals can undercover (but not the extreme as DML, Spatial+), we consider the use of selection criteria to determine whether a variable is important. We compute the energy score, where recall from Proposition 1 and Section 2.5, is identical between the SLMM and the RSR. The energy score (scaled so that smaller values are preferred) is 851.54 for the model that excludes PM2.5 and 779.97 for the model that includes it, which coincides with the conclusions from the RSR and the adjusted RSR centered on the OLS estimator.

\begin{figure}
    \centering
    \includegraphics{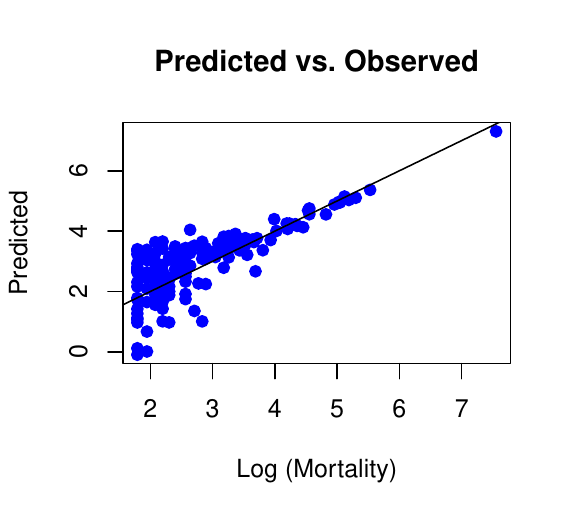}
    \caption{Predicted values via the aRSR versus the observed mortality on the log-scale.}
    \label{fig:pred:aRSR}
\end{figure}

\section{Discussion} \label{sec:conc}

In this article, we show that traditional RSRs produce identical inferences for missing values, $\bm{\delta}$, and hyperparameters as Bayesian SLMMs (with improper prior on $\bm{\beta}$) even when the data is generated from the SLMM. Thus, the RSR is ineffectual in terms of statistical inference on $\bm{\delta}$ and missing data as compared to the SLMM under certain prior assumptions. It is also shown that there exists an augmented RSR (the same considered in \citet{hanks2015restricted}) that produces the same inferences on linearly dependent regression coefficients (upon a posterior predictive adjustment), spatial random effects, missing values, and hyperparameters as the SLMM. These results lead to several equivalence relationships that the Bayes factor, strictly proper scoring rules, BMA, and out-of-sample predictive metrics all measure the SLMM and RSR as equivalent, which indicates that both are equally reasonable models to consider when adopting a formal model selection perspective.

Our conclusions differ from that of \citet{zimmerman2022deconfounding} and \cite{khan2022restricted} because we take a traditional model selection view for comparing two models, where neither model is assumed to be necessarily true. Conclusions 2 and 3 suggest that the misspecified RSR is generally equivalent to the SLMM under several selection criteria provided one implements the RSR in a Bayesian setting with location invariant priors. However, this perspective does not resolve the fact that the misspecified RSR is estimating a different quantity than the SLMM (i.e., orthogonal regression effects versus linearly dependent regression effects). Thus, we propose a kind of transfer learning strategy similar to \citet{hanks2015restricted} to adjust for this difference in a manner that aims at obtaining robust estimates in the misspecified setting. As such, these results motivate a recommendation to adopt a Bayesian framework when using an RSR, location invariant prior distributions, our posterior predictive strategy to estimate linearly dependent regression effects, and the use of selection criteria to determine important covariates.

The RSR and aRSR aim to provide a solution for unmeasured confounders via robustness properties, and does not address other important assumptions needed to draw causal assumptions including positivity, SUTVA, and consistency that are prevalent in the potential outcome framework for causal analyses \citep{rubin2005causal}. Consequently, additional strategies are needed when one suspects that any of these assumptions fail \citep[e.g., see][for a review of such strategies]{reich2021review}.

Algorithm 1 shows that there is little computational overhead involved with sampling linearly dependent regression effects from the SLMM and the versions of aRSR considered in this paper (i.e., Steps  7, 8, 9 in Algorithm 1). Consequently, in practice, it is fairly straightforward to consider each estimator of the linearly dependent regression effect. If it is expected that unmeasured confounders are present then the results in this article suggest that $\bm{\beta}^{adj}_{trn}$ is a reasonable option to consider. 

In our simulations, we consider a more standard simulation setup similar to that in \citet{paciorek2010importance}, modified to allow for random $\bm{\beta}$, where we see that our transfer learning strategy and the traditional RSR can lead to improvements in root mean squared error, prediction, and coverage as compared to the SLMM, Spatial+, and the DML shift estimator. In our application to COVID-19 mortality we were able to detect a significant effect of PM2.5 via inference on $\bm{\delta}$ and our transfer learning strategy on linearly dependent regression effects, but failed to detect a significant effect of PM2.5 when using the SLMM. The energy score metric implied that PM2.5 is a useful covariate aligning more with RSR (restricted to either $\bm{\beta} = \delta$ or $\bm{\beta} = \bm{\beta}_{trn}^{adj}$) than the SLMM.

\bibliographystyle{jasa}

\bibliography{myref33}
%

\section*{Appendix A: Technical Results}

\noindent
\textit{Proof of Proposition 2:} After applying the change of variables in (\ref{eq:changeofvariables}) to the hierarchical model in (\ref{eq:parametric}), it is enough to show that $f(\bm{\delta}\vert \bm{\nu} ,\textbf{y},\textbf{X},\bm{\theta}) = N\left\lbrace(\textbf{X}^{\prime}\textbf{X})^{-1}\textbf{X}^{\prime}\textbf{y},\sigma^{2}(\textbf{X}^{\prime}\textbf{X})^{-1}\right\rbrace$. {To prove Equation (\ref{eq:OLS}), we have that}
\begin{align*}
	&f(\bm{\delta}\vert \bm{\nu} ,\textbf{y},\textbf{X},\textbf{Z},{\bm{\theta}})\propto f(\textbf{y}\vert \textbf{X}, \bm{\delta},\bm{\nu},\sigma^{2})\\
	& \propto\mathrm{exp}\left[-\frac{1}{2\sigma^{2}}\left\lbrace\textbf{y} - \textbf{X}\bm{\delta} - (\textbf{I}-\textbf{P})\bm{\nu}\right\rbrace^{\prime}\left\lbrace\textbf{y} - \textbf{X}\bm{\delta} - (\textbf{I}-\textbf{P})\bm{\nu}\right\rbrace\right]\\
	& \propto\mathrm{exp}\left[-\frac{1}{2}\left\lbrace\bm{\delta}^{\prime}\left(\frac{1}{\sigma^{2}}\textbf{X}^{\prime}\textbf{X}\right)\bm{\delta} - 2\bm{\delta}^{\prime}\textbf{X}^{\prime}\left(\frac{1}{\sigma^{2}}\textbf{y} - \frac{1}{\sigma^{2}}(\textbf{I}-\textbf{P})\bm{\nu}\right)\right\rbrace\right]\\
	& =\mathrm{exp}\left[-\frac{1}{2}\left\lbrace\bm{\delta}^{\prime}\left(\frac{1}{\sigma^{2}}\textbf{X}^{\prime}\textbf{X}\right)\bm{\delta} - 2\bm{\delta}^{\prime}\textbf{X}^{\prime}\left(\frac{1}{\sigma^{2}}\textbf{y}\right)\right\rbrace\right],
\end{align*}
\noindent
and we have the result upon multiplying by $\left(\frac{1}{\sigma^{2}}\textbf{X}^{\prime}\textbf{X}\right)\left(\frac{1}{\sigma^{2}}\textbf{X}^{\prime}\textbf{X}\right)^{-1}$ in the cross-product term and completing the squares. {Notice that 
\begin{align*}
	f(\bm{\delta}\vert \textbf{y},\textbf{X},\sigma^{2}) &= \int \int f(\bm{\delta}\vert \bm{\nu} ,\textbf{y},\textbf{X},\textbf{Z},{\bm{\theta}}) f(\bm{\nu},\bm{\theta}_{-\sigma^{2}}\vert \textbf{y},\textbf{X},\textbf{Z},\sigma^{2}) d\bm{\theta}_{-\sigma^{2}} d\bm{\nu}\\
	&=f(\bm{\delta}\vert \bm{\nu} ,\textbf{y},\textbf{X},\textbf{Z},{\bm{\theta}})\int \int  f(\bm{\nu},\bm{\theta}_{-\sigma^{2}}\vert \textbf{y},\textbf{X},\textbf{Z},\sigma^{2}) d\bm{\theta}_{-\sigma^{2}} d\bm{\nu}\\
	&=f(\bm{\delta}\vert \bm{\nu} ,\textbf{y},\textbf{X},\textbf{Z},{\bm{\theta}}),
\end{align*}
where $\bm{\theta}_{-\sigma^{2}}$ contains all elements of $\bm{\theta}$ except $\sigma^{2}$ and we can pull $f(\bm{\delta}\vert \bm{\nu} ,\textbf{y},\textbf{X},\textbf{Z},{\bm{\theta}})$ outside the integral, since from the argument above, $f(\bm{\delta}\vert \bm{\nu} ,\textbf{y},\textbf{X},\textbf{Z},{\bm{\theta}})$ does not contain $\bm{\nu}$ and $\bm{\theta}_{-\sigma^{2}}$.}

{{In a similar manner, to prove Equation (\ref{eq:OLS2}), we have that}
\begin{align*}
	&f(\bm{\delta}\vert \bm{\nu} ,\textbf{y},\textbf{X},{\bm{\theta}})\propto f(\textbf{y}\vert \textbf{X}, \bm{\delta},\bm{\nu},\sigma^{2})\\
	& \propto\mathrm{exp}\left[-\frac{1}{2\sigma^{2}}\left\lbrace\textbf{y} - \textbf{X}\bm{\delta} - (\textbf{I}-\textbf{P})\bm{\nu}\right\rbrace^{\prime}\left\lbrace\textbf{y} - \textbf{X}\bm{\delta} - (\textbf{I}-\textbf{P})\bm{\nu}\right\rbrace\right]\\
	& \propto\mathrm{exp}\left[-\frac{1}{2}\left\lbrace\bm{\delta}^{\prime}\left(\frac{1}{\sigma^{2}}\textbf{X}^{\prime}\textbf{X}\right)\bm{\delta} - 2\bm{\delta}^{\prime}\textbf{X}^{\prime}\left(\frac{1}{\sigma^{2}}\textbf{y} - \frac{1}{\sigma^{2}}(\textbf{I}-\textbf{P})\bm{\nu}\right)\right\rbrace\right]\\
	& =\mathrm{exp}\left[-\frac{1}{2}\left\lbrace\bm{\delta}^{\prime}\left(\frac{1}{\sigma^{2}}\textbf{X}^{\prime}\textbf{X}\right)\bm{\delta} - 2\bm{\delta}^{\prime}\textbf{X}^{\prime}\left(\frac{1}{\sigma^{2}}\textbf{y}\right)\right\rbrace\right],
\end{align*}
	\noindent
	and we have the result upon multiplying by $\left(\frac{1}{\sigma^{2}}\textbf{X}^{\prime}\textbf{X}\right)\left(\frac{1}{\sigma^{2}}\textbf{X}^{\prime}\textbf{X}\right)^{-1}$ in the cross-product term and completing the squares. {Notice that 
		\begin{align*}
			f(\bm{\delta}\vert \textbf{y},\textbf{X},\sigma^{2}) &=  \int f(\bm{\delta}\vert \textbf{y},\textbf{X},\bm{\nu},\sigma^{2}) f(\bm{\nu}\vert \textbf{y},\textbf{X},\sigma^{2})  d\bm{\nu}\\
			&=f(\bm{\delta}\vert \textbf{y},\textbf{X},\bm{\nu},\sigma^{2}) \int  f(\bm{\nu}\vert \textbf{y},\textbf{X},\sigma^{2})  d\bm{\nu}\\
			&=f(\bm{\delta}\vert \textbf{y},\textbf{X},\bm{\nu},\sigma^{2}),
		\end{align*}
		where we can pull $f(\bm{\delta}\vert \textbf{y},\textbf{X},\bm{\nu},\sigma^{2})$ outside the integral, since from the argument above, $f(\bm{\delta}\vert \textbf{y},\textbf{X},\bm{\nu},\sigma^{2})$ does not contain $\bm{\nu}$.}}\\

\noindent
\textbf{Formal Statement of Proposition 3:} Assume the hierarchical model in (\ref{eq:rsr:bhm:aug2}). Then the posterior distribution can be decomposed according to (\ref{eq:joint:post}) such that,
\begin{align}
	\nonumber
	&f_{trn}(\bm{\beta}^{adj}_{trn}\vert \bm{\beta}_{RSR},\sigma^{2}\bm{\Sigma}_{\nu}(\bm{\Lambda}),\bm{\mu}_{\beta},\textbf{X})\\
	\nonumber
	&f_{aRSR}(\bm{\beta}_{RSR}\vert \textbf{y},\sigma^{2},\textbf{X}) = N\left\lbrace (\textbf{X}^{\prime}\textbf{X})^{-1}\textbf{X}^{\prime}\textbf{y},\sigma^{2}(\textbf{X}^{\prime}\textbf{X})^{-1}\right\rbrace\\
	\nonumber
	&f_{aRSR}(\bm{\nu}\vert \textbf{y},\sigma^{2},\bm{\Lambda},\textbf{X}) = N\left\lbrace \left((\textbf{I}-\textbf{P})+\bm{\Sigma}_{\nu}^{-1}\right)^{-1}(\textbf{I}-\textbf{P})\textbf{y},\sigma^{2}\left((\textbf{I}-\textbf{P})+\bm{\Sigma}_{\nu}^{-1}\right)^{-1}\right\rbrace\\
	\nonumber
	&f_{aRSR}(\sigma^{2}\vert \textbf{y},\bm{\Lambda},\textbf{X}) = IG(\alpha^{*},\kappa^{*})\\
	\nonumber
	&f_{aRSR}(\textbf{y}_{m}\vert \textbf{y}_{o},\bm{\Lambda},\textbf{X})= \mathcal{MT}(\bm{\Sigma}_{m,o}\bm{\Sigma}_{o}^{-1}\textbf{y}_{o},\rho(\bm{\Sigma}_{m}-\bm{\Sigma}_{m,o}\bm{\Sigma}_{o}^{-1}\bm{\Sigma}_{o,m}), 2\alpha + n_{o}-p)\\
	\label{eq:post}
	&f(\mathrm{vec}(\bm{\Lambda})\vert \textbf{y}_{o},\textbf{X}) = \frac{\Gamma(\frac{2\alpha-p+n_{o}}{2})}{\Gamma((2\alpha-p)/2)\gamma^{n_{o}/2}(\prod_{i = 1}^{n_{o}}(1/h_{i}^{2}))^{1/2}} \frac{1}{\left(\frac{1}{2\alpha-p}\sum_{k = 1}^{n_{o}}\frac{h_{k}^{2}}{\lambda_{k}+1}+1\right)^{\frac{n_{o}+2\alpha-p}{2}}} \prod_{i = 1}^{n_{o}}\left(\frac{1}{\lambda_{i}+1}\right)^{3/2},
\end{align}
\noindent
where the function ``vec$(\bm{\Lambda})$'' produces the vector along the main diagonal of $\bm{\Lambda}$, $\alpha^{*} = (n-p)/2+\alpha$, $\kappa^{*} = \textbf{y}^{\prime}\left((\textbf{I}-\textbf{P})\bm{\Sigma}_{\nu}(\textbf{I}-\textbf{P})+\textbf{I}\right)^{-1}\textbf{y}/2 + \kappa$, $\bm{\Sigma}_{n} = \frac{2\kappa}{(2\alpha-p)} \left((\textbf{I}-\textbf{P})\bm{\Sigma}_{\nu}(\textbf{I}-\textbf{P})+\textbf{I}\right)$, $\bm{\Sigma}_{o} = (\bm{0}_{n_{o},n_{m}},\textbf{I}_{n_{o}})\bm{\Sigma}_{n}(\bm{0}_{n_{o},n_{m}},\textbf{I}_{n_{o}})^{\prime}$, $\bm{\Sigma}_{m} = (\textbf{I}_{n_{m}},\bm{0}_{n_{m},n_{o}})\bm{\Sigma}_{n}(\textbf{I}_{n_{m}},\bm{0}_{n_{m},n_{o}})^{\prime}$, \\ $\bm{\Sigma}_{m,o} = (\textbf{I}_{n_{m}},\bm{0}_{n_{m},n_{o}})\bm{\Sigma}_{n}(\bm{0}_{n_{o},n_{m}},\textbf{I}_{n_{o}})^{\prime}$, $\bm{\Sigma}_{o,m} = (\bm{0}_{n_{o},n_{m}},\textbf{I}_{n_{o}})\bm{\Sigma}_{n}(\textbf{I}_{n_{m}},\bm{0}_{n_{m},n_{o}})^{\prime}$, $\rho = \frac{(2\alpha-p) + \textbf{y}_{o}^{\prime}\bm{\Sigma}_{o}^{-1}\textbf{y}_{o}}{(2\alpha+n_{o}-p)}$, $\textbf{h} = (h_{1},\ldots,h_{n_{o}})^{\prime} = \left(\frac{2\alpha - p}{2\kappa}\right)^{1/2}\textbf{B}^{\prime}\textbf{y}_{o}$, and $\mathcal{MT}(\bm{\mu},\textbf{C},w)$ is a shorthand for the multivariate $t$ distribution with real-vector-valued mean $\bm{\mu}$, positive definite covariance matrix $\textbf{C}$, and degrees of freedom $w>0$. \\

\noindent
\textit{Proof of $f_{trn}(\bm{\beta}^{adj}_{trn}\vert \bm{\beta}_{RSR},\sigma^{2}\bm{\Sigma}_{\nu}(\bm{\Lambda}),\bm{\mu}_{\beta})$  and $f_{aRSR}(\bm{\beta}_{RSR}\vert \textbf{y},\sigma^{2})$ in (\ref{eq:post}):} The term\\ $f_{trn}(\bm{\beta}^{adj}_{trn}\vert \bm{\beta}_{RSR},\sigma^{2}\bm{\Sigma}_{\nu}(\bm{\Lambda}),\bm{\mu}_{\beta})$ follows by definition, and $f_{aRSR}(\bm{\beta}_{RSR}\vert \textbf{y},\sigma^{2})$ follows from Proposition 2.\\

\noindent
\textit{Proof of the Expression of $f_{aRSR}(\bm{\nu}\vert \textbf{y},\textbf{X},\sigma^{2},\bm{\Lambda})$ in (\ref{eq:post}):}  The goal is to show that
\begin{align}\label{eq:nu:fullcond}
	&f(\bm{\nu}\vert \textbf{y},\textbf{X},\bm{\delta},\bm{\theta}) \\
	\nonumber
	&=N\left\lbrace \frac{1}{\sigma^{2}}\left(\frac{1}{\sigma^{2}}(\textbf{I}-\textbf{P})+\frac{1}{\sigma^{2}}(\bm{\Sigma}_{\nu})^{-1}\right)^{-1} (\textbf{I}-\textbf{P})\textbf{y},\left(\frac{1}{\sigma^{2}}(\textbf{I}-\textbf{P})+\frac{1}{\sigma^{2}}(\bm{\Sigma}_{\nu})^{-1}\right)^{-1}\right\rbrace.
\end{align}
\noindent
We have the following proportionality argument,
\begin{align*}
	&f(\bm{\nu}\vert \textbf{y},\textbf{X},\bm{\delta},\bm{\theta})\propto f(\textbf{y}\vert \textbf{X}, \bm{\delta},\bm{\nu},\sigma^{2})f(\bm{\nu} \vert \textbf{X}, \bm{\theta})\\
	& \propto\mathrm{exp}\left[-\frac{1}{2\sigma^{2}}\left\lbrace\textbf{y} - \textbf{X}\bm{\delta} - (\textbf{I}-\textbf{P})\bm{\nu}\right\rbrace^{\prime}\left\lbrace\textbf{y} - \textbf{X}\bm{\delta} - (\textbf{I}-\textbf{P})\bm{\nu}\right\rbrace\right.\\
	&\left. \hspace{100pt}- \frac{1}{2\sigma^{2}}\bm{\nu}^{\prime}(\bm{\Sigma}_{\nu})^{-1}\bm{\nu}\right]\\
	& \propto\mathrm{exp}\left[-\frac{1}{2}\left\lbrace \bm{\nu}^{\prime}\left(\frac{1}{\sigma^{2}}(\textbf{I}-\textbf{P})+\frac{1}{\sigma^{2}}(\bm{\Sigma}_{\nu})^{-1}\right)\bm{\nu}\right.\right.\\
	&\hspace{100pt}\left.\left. - 2\bm{\nu}^{\prime}(\textbf{I}-\textbf{P})\left(\frac{1}{\sigma^{2}}\textbf{y} - \frac{1}{\sigma^{2}}\textbf{X}\bm{\delta}\right)  \right\rbrace\right]\\
	& \propto\mathrm{exp}\left[-\frac{1}{2}\left\lbrace \bm{\nu}^{\prime}\left(\frac{1}{\sigma^{2}}(\textbf{I}-\textbf{P})+\frac{1}{\sigma^{2}}(\bm{\Sigma}_{\nu})^{-1}\right)\bm{\nu} - 2\bm{\nu}^{\prime}(\textbf{I}-\textbf{P})\left(\frac{1}{\sigma^{2}}\textbf{y}\right)  \right\rbrace\right],
\end{align*}
\noindent
and we have (\ref{eq:nu:fullcond}) upon multiplying by $\left(\frac{1}{\sigma^{2}}(\textbf{I}-\textbf{P})+\frac{1}{\sigma^{2}}(\bm{\Sigma}_{\nu})^{-1}\right)\left(\frac{1}{\sigma^{2}}(\textbf{I}-\textbf{P})+\frac{1}{\sigma^{2}}(\bm{\Sigma}_{\nu})^{-1}\right)^{-1}$ in the cross-product term and completing the squares. {Notice that 
	\begin{align*}
		f(\bm{\nu}\vert  \textbf{y},\textbf{X},\bm{\theta}) &=  \int f(\bm{\nu}\vert  \textbf{y},\textbf{X},\bm{\delta},\bm{\theta}) f(\bm{\delta}\vert \textbf{y},\textbf{X},\bm{\theta}) d\bm{\delta}\\
		&=f(\bm{\nu}\vert  \textbf{y},\textbf{X},\bm{\delta},\bm{\theta})\int f(\bm{\delta}\vert \textbf{y},\textbf{X},\bm{\theta}) d\bm{\delta}\\
		&=f(\bm{\nu}\vert  \textbf{y},\textbf{X},\bm{\delta},\bm{\theta}),
	\end{align*}
	where we can pull $f(\bm{\nu}\vert  \textbf{y},\textbf{X},\bm{\delta},\bm{\theta})$ outside the integral, since from the argument above, $f(\bm{\nu}\vert  \textbf{y},\textbf{X},\bm{\delta},\bm{\theta})$ does not contain $\bm{\delta}$. Consequently, $\bm{\nu}$ is conditionally independent of $\bm{\delta}$ given $\textbf{y}$, $\textbf{X}$, and $\bm{\theta}$.}\\

\noindent
\textit{Proof of the expression of $f_{aRSR}(\sigma^{2}\vert \textbf{y},\textbf{X},\bm{\Lambda})$ in (\ref{eq:post}):} Let $\bm{\theta} = \{\sigma^{2},\bm{\Lambda}\}$. Note that from the result above, we have
\begin{equation}\label{eq:initial}
	f(\bm{\nu}\vert \textbf{y},\textbf{X},\bm{\theta}) = \frac{f(\bm{\nu}, \textbf{y},\textbf{X},\bm{\theta})}{f( \textbf{y},\textbf{X},\bm{\theta})},
\end{equation}
where the left-hand-side of (\ref{eq:initial}) is given by,
\begin{align}\nonumber
	&f(\bm{\nu}\vert \textbf{y},\textbf{X},\bm{\theta}) = N\left\lbrace \frac{1}{\sigma^{2}}\left(\frac{1}{\sigma^{2}}(\textbf{I}-\textbf{P})+\frac{1}{\sigma^{2}}(\bm{\Sigma}_{\nu})^{-1}\right)^{-1} (\textbf{I}-\textbf{P})\textbf{y},\left(\frac{1}{\sigma^{2}}(\textbf{I}-\textbf{P})+\frac{1}{\sigma^{2}}(\bm{\Sigma}_{\nu})^{-1}\right)^{-1}\right\rbrace\\
	\nonumber
	&=\frac{1}{\mathcal{N}_{1}} exp\left\lbrace -\frac{1}{2}\left(\bm{\nu} - \frac{1}{\sigma^{2}}\left(\frac{1}{\sigma^{2}}(\textbf{I}-\textbf{P})+\frac{1}{\sigma^{2}}(\bm{\Sigma}_{\nu})^{-1}\right)^{-1} (\textbf{I}-\textbf{P})\textbf{y}\right)^{\prime}\right.\\
	\nonumber
	&\left.\left(\frac{1}{\sigma^{2}}(\textbf{I}-\textbf{P})+\frac{1}{\sigma^{2}}(\bm{\Sigma}_{\nu})^{-1}\right)\left(\bm{\nu} - \frac{1}{\sigma^{2}}\left(\frac{1}{\sigma^{2}}(\textbf{I}-\textbf{P})+\frac{1}{\sigma^{2}}(\bm{\Sigma}_{\nu})^{-1}\right)^{-1} (\textbf{I}-\textbf{P})\textbf{y}\right)\right\rbrace\\
	\nonumber
	&= \frac{1}{\mathcal{N}_{1}}exp\left(-\frac{1}{2\sigma^{2}}\bm{\nu}^{\prime}(\textbf{I}-\textbf{P})\bm{\nu}-\frac{1}{2\sigma^{2}}\bm{\nu}^{\prime}(\bm{\Sigma}_{\nu})^{-1}\bm{\nu}\right.\\
	\label{eq:second}
	&\left.+\frac{1}{\sigma^{2}}\bm{\nu}^{\prime}(\textbf{I}-\textbf{P})\textbf{y}-\frac{1}{2}\left(\frac{1}{\sigma^{2}}\right)^{2}\textbf{y}^{\prime}(\textbf{I}-\textbf{P})\left(\frac{1}{\sigma^{2}}(\textbf{I}-\textbf{P})+\frac{1}{\sigma^{2}}(\bm{\Sigma}_{\nu})^{-1}\right)^{-1}(\textbf{I}-\textbf{P})\textbf{y}\right),
\end{align}
where $1/\mathcal{N}_{1}$ is the normalizing constant. We also have
\begin{align}
	\nonumber
	&f(\textbf{y},\textbf{X},\bm{\nu},\bm{\theta}) = f(\textbf{y}\vert  \textbf{X},\bm{\nu},\bm{\theta}) f(\bm{\nu}\vert \textbf{X},\bm{\theta}) f(\bm{\theta})\\
	\nonumber
	& = \int f(\textbf{y}\vert \textbf{X},\bm{\delta},\bm{\nu}, \bm{\theta}) f(\bm{\delta}\vert \textbf{X},\bm{\nu},\bm{\theta}) d\bm{\delta} f(\bm{\nu}\vert \textbf{X},\bm{\theta}) f(\bm{\theta})\\
	\nonumber
	& = \int \frac{\mathrm{exp}\left[-\frac{1}{2\sigma^{2}}\left\lbrace\textbf{y} - \textbf{X}\bm{\delta} - (\textbf{I}-\textbf{P})\bm{\nu}\right\rbrace^{\prime}\left\lbrace\textbf{y} - \textbf{X}\bm{\delta} - (\textbf{I}-\textbf{P})\bm{\nu}\right\rbrace \right]}{\mathcal{N}_{2}}
	d\bm{\delta}\\
	\nonumber
	& f(\bm{\nu}\vert \textbf{X},\bm{\theta}) f(\bm{\theta}),
\end{align}
\noindent
where $1/\mathcal{N}_{2}$ is the normalizing constant for $f(\textbf{y}\vert \textbf{X},\bm{\delta},\bm{\nu}, \textbf{X},\bm{\theta})$ and recall $f(\bm{\delta}\vert \textbf{X},\bm{\nu},\bm{\theta}) = 1$, so that we have that the above,
\begin{align}
	\nonumber
	& = \int \frac{\mathrm{exp}\left[-\frac{1}{2\sigma^{2}}\left\lbrace\textbf{y} - \textbf{X}\bm{\delta} - (\textbf{I}-\textbf{P})\bm{\nu}\right\rbrace^{\prime}\left\lbrace\textbf{y} - \textbf{X}\bm{\delta} - (\textbf{I}-\textbf{P})\bm{\nu}\right\rbrace\right]}{\mathcal{N}_{2}}
	d\bm{\delta}\\
	\nonumber
	& \frac{1}{\mathcal{N}_{3}}\mathrm{exp}\left(- \frac{1}{2\sigma^{2}}\bm{\nu}^{\prime}(\bm{\Sigma}_{\nu})^{-1}\bm{\nu}\right) f(\bm{\theta})\\
	\nonumber
	& = \frac{\mathrm{exp}\left[-\frac{1}{2\sigma^{2}}\left\lbrace\textbf{y} - (\textbf{I}-\textbf{P})\bm{\nu}\right\rbrace^{\prime}\left\lbrace\textbf{y} - (\textbf{I}-\textbf{P})\bm{\nu}\right\rbrace\right]}{\mathcal{N}_{2}}\\
	\nonumber
	&\int \mathrm{exp}\left[-\frac{1}{2\sigma^{2}}\bm{\delta}^{\prime}(\textbf{X}^{\prime}\textbf{X})\bm{\delta} + \frac{1}{\sigma^{2}}\bm{\delta}^{\prime}\textbf{X}\textbf{y}\right]
	d\bm{\delta}\\
	\nonumber
	& \frac{1}{\mathcal{N}_{3}}\mathrm{exp}\left(- \frac{1}{2\sigma^{2}}\bm{\nu}^{\prime}(\bm{\Sigma}_{\nu})^{-1}\bm{\nu}\right) f(\bm{\theta})\\
	\nonumber
	& = \frac{\mathrm{exp}\left[-\frac{1}{2\sigma^{2}}\textbf{y}^{\prime}\textbf{y}-\frac{1}{2\sigma^{2}}\bm{\nu}^{\prime}(\textbf{I}-\textbf{P})\bm{\nu} +\frac{1}{\sigma^{2}}\bm{\nu}^{\prime}(\textbf{I}-\textbf{P})\textbf{y}\right]}{\mathcal{N}_{2}}\\
	\nonumber
	&\mathcal{N}_{4}{\mathrm{exp}\left(\frac{1}{\sigma^{2}}\textbf{y}^{\prime}\textbf{P}\textbf{y}\right)}\int N\left\lbrace(\textbf{X}^{\prime}\textbf{X})^{-1}\textbf{X}^{\prime}\textbf{y},\sigma^{2}(\textbf{X}^{\prime}\textbf{X})^{-1}\right\rbrace
	d\bm{\delta}\\
	\label{eq:third}
	& \frac{1}{\mathcal{N}_{3}}\mathrm{exp}\left(- \frac{1}{2\sigma^{2}}\bm{\nu}^{\prime}(\bm{\Sigma}_{\nu})^{-1}\bm{\nu}\right) f(\bm{\theta}),
\end{align}
\noindent
with $1/\mathcal{N}_{3}$ the normalizing constant for $f(\bm{\nu}\vert \textbf{X},\bm{\theta})$ and $1/\mathcal{N}_{4}$ the normalizing constant for $ N\left\lbrace(\textbf{X}^{\prime}\textbf{X})^{-1}\textbf{X}^{\prime}\textbf{y},\sigma^{2}(\textbf{X}^{\prime}\textbf{X})^{-1}\right\rbrace$. Substituting (\ref{eq:second}) and (\ref{eq:third}) into (\ref{eq:initial}), we have
\begin{align}
	\nonumber\label{eq:marg}
	&f(\textbf{y},\textbf{X},\bm{\theta})=\frac{\mathcal{N}_{1}\mathcal{N}_{4}}{\mathcal{N}_{2}\mathcal{N}_{3}}\\
	&\times exp\left\lbrace-\frac{\textbf{y}^{\prime}(\textbf{I}-\textbf{P})\textbf{y}/2 + \textbf{y}^{\prime}\textbf{P}\textbf{y}/2}{\sigma^{2}}+\frac{1}{2}\left(\frac{1}{\sigma^{2}}\right)^{2}\textbf{y}^{\prime}(\textbf{I}-\textbf{P})\left(\frac{1}{\sigma^{2}}(\textbf{I}-\textbf{P})+\frac{1}{\sigma^{2}}(\bm{\Sigma}_{\nu})^{-1}\right)^{-1}(\textbf{I}-\textbf{P})\textbf{y}\right\rbrace f(\bm{\theta}),
\end{align}
\noindent
where,
\begin{align*}
	\mathcal{N}_{1} & = (2\pi)^{n/2}det\left\lbrace \left(\frac{1}{\sigma^{2}}(\textbf{I}-\textbf{P})+\frac{1}{\sigma^{2}}(\bm{\Sigma}_{\nu})^{-1}\right)^{-1}\right\rbrace^{1/2}\\
	\mathcal{N}_{2} & = (2\pi)^{n/2}\left\lbrace\sigma^{2}\right\rbrace^{n/2}\\
	\mathcal{N}_{3} & = (2\pi)^{n/2}det\left(\sigma^{2}\bm{\Sigma}_{\nu}\right)^{1/2}\\
	\mathcal{N}_{4} & = (2\pi)^{p/2}det\left\lbrace \sigma^{2}\left(\textbf{X}^{\prime}\textbf{X}\right)^{-1}\right\rbrace^{1/2}.
\end{align*}
\noindent
It follows that
\begin{align}\label{eq:marg:sig}
	\nonumber
	&f(\sigma^{2}\vert \textbf{y},\textbf{X},\bm{\Lambda})\\
	\nonumber
	&\propto (\sigma^{2})^{-(n-p)/2-\alpha-1} exp\left\lbrace -\frac{\textbf{y}^{\prime}(\textbf{I}-\textbf{P})\textbf{y}/2 + \textbf{y}^{\prime}\textbf{P}\textbf{y}/2 - \textbf{y}^{\prime}(\textbf{I}-\textbf{P})\left((\textbf{I}-\textbf{P})+(\bm{\Sigma}_{\nu})^{-1}\right)^{-1}(\textbf{I}-\textbf{P})\textbf{y}/2 + \kappa}{\sigma^{2}}\right\rbrace\\
	&\propto IG((n-p)/2+\alpha, \textbf{y}^{\prime}(\textbf{I}-\textbf{P})\textbf{y}/2 + \textbf{y}^{\prime}\textbf{P}\textbf{y}/2 - \textbf{y}^{\prime}(\textbf{I}-\textbf{P})\left((\textbf{I}-\textbf{P})+(\bm{\Sigma}_{\nu})^{-1}\right)^{-1}(\textbf{I}-\textbf{P})\textbf{y}/2 + \kappa).
\end{align}
\noindent
To simplify $\kappa^{*}$ we have,
\begin{align}
	\nonumber
	\kappa^{*}-\kappa=&\textbf{y}^{\prime}(\textbf{I}-\textbf{P})\textbf{y} - \textbf{y}^{\prime}(\textbf{I}-\textbf{P})\left((\textbf{I}-\textbf{P})+(\bm{\Sigma}_{\nu})^{-1}\right)^{-1}(\textbf{I}-\textbf{P})\textbf{y} + \textbf{y}^{\prime}\textbf{P}\textbf{y}/2\\
	\nonumber
	& = \textbf{y}^{\prime}(\textbf{I}-\textbf{P})(\textbf{I}-\textbf{P})\textbf{y} -\textbf{y}^{\prime}(\textbf{I}-\textbf{P})\left((\textbf{I}-\textbf{P})+(\bm{\Sigma}_{\nu})^{-1}\right)^{-1}(\textbf{I}-\textbf{P})\textbf{y} + \textbf{y}^{\prime}\textbf{P}\textbf{y}/2\\
	\nonumber
	&= \textbf{y}^{\prime}(\textbf{I}-\textbf{P})\left(\textbf{I} - (\textbf{I}-\textbf{P})\left((\textbf{I}-\textbf{P})(\textbf{I}-\textbf{P})+(\bm{\Sigma}_{\nu})^{-1}\right)^{-1}(\textbf{I}-\textbf{P})\right)(\textbf{I}-\textbf{P})\textbf{y} + \textbf{y}^{\prime}\textbf{P}\textbf{y}/2\\
	\label{eq:sherman}
	&= \textbf{y}^{\prime}(\textbf{I}-\textbf{P})\left((\textbf{I}-\textbf{P})\bm{\Sigma}_{\nu}(\textbf{I}-\textbf{P})+\textbf{I}\right)^{-1}(\textbf{I}-\textbf{P})\textbf{y} + \textbf{y}^{\prime}\textbf{P}\textbf{y}/2,
\end{align} 
\noindent
where the last equality holds by the Sherman-Morrison-Woodbury formula \citep{johan}. Using the Neumann series expression,
\begin{align*}
	\kappa^{*}-\kappa &= \textbf{y}^{\prime}(\textbf{I}-\textbf{P})\left((\textbf{I}-\textbf{P})\bm{\Sigma}_{\nu}(\textbf{I}-\textbf{P})+\textbf{I}\right)^{-1}(\textbf{I}-\textbf{P})\textbf{y}/2 + \textbf{y}^{\prime}\textbf{P}\textbf{y}/2\\
	&= \textbf{y}^{\prime}(\textbf{I}-\textbf{P})\left\lbrace\sum_{k = 0}^{\infty}\left(-(\textbf{I}-\textbf{P})\bm{\Sigma}_{\nu}(\textbf{I}-\textbf{P})\right)^{k}\right\rbrace (\textbf{I}-\textbf{P})\textbf{y}/2+ \textbf{y}^{\prime}\textbf{P}\textbf{y}/2\\
	&= \textbf{y}^{\prime}\left\lbrace(\textbf{I}-\textbf{P})+ \sum_{k = 1}^{\infty}\left(-(\textbf{I}-\textbf{P})\bm{\Sigma}_{\nu}(\textbf{I}-\textbf{P})\right)^{k}\right\rbrace \textbf{y}/2+ \textbf{y}^{\prime}\textbf{P}\textbf{y}/2\\
		&= \textbf{y}^{\prime}\left\lbrace \sum_{k = 0}^{\infty}\left((\textbf{I}-\textbf{P})\bm{\Sigma}_{\nu}(\textbf{I}-\textbf{P})\right)^{k}-\textbf{P}\right\rbrace \textbf{y}/2+ \textbf{y}^{\prime}\textbf{P}\textbf{y}/2\\
		&= \textbf{y}^{\prime}\left((\textbf{I}-\textbf{P})\bm{\Sigma}_{\nu}(\textbf{I}-\textbf{P})+\textbf{I}\right)^{-1}\textbf{y}/2,\\
\end{align*}
where the third equality follows from the dact that $(\textbf{I}-\textbf{P})$ is idempotent. Moreover, the above expression is strictly positive because $\left((\textbf{I}-\textbf{P})\bm{\Sigma}_{\nu}(\textbf{I}-\textbf{P})+\textbf{I}\right)^{-1}$ is positive definite. 

\noindent
\textit{Proof of the expression of $f_{aRSR}(\textbf{y}_{m}\vert \textbf{y}_{o},\bm{\Lambda})$ in (\ref{eq:post}):} It follows from the fact that (\ref{eq:marg}) divided by (\ref{eq:marg:sig}) gives
\begin{align*}
	f(\textbf{y},\textbf{X},\bm{\Lambda}) &= \frac{det\left\lbrace(\textbf{X}^{\prime}\textbf{X})^{-1}\right\rbrace^{1/2}det\left\lbrace \left((\textbf{I}-\textbf{P})+(\bm{\Sigma}_{\nu})^{-1}\right)^{-1}\right\rbrace^{1/2} det((\bm{\Sigma}_{\nu})^{-1})^{1/2} f(\bm{\Lambda})}{(2\pi)^{(n-p)/2} (\textbf{y}^{\prime}\left((\textbf{I}-\textbf{P})\bm{\Sigma}_{\nu}(\textbf{I}-\textbf{P})+\textbf{I}\right)^{-1}\textbf{y}/2+ \kappa)^{(n-p)/2+\alpha}}.
\end{align*}
\noindent
and using the Woodbury determinant lemma,
\begin{align}
	\label{eq:marg:post:propto}
	f(\textbf{y},\textbf{X},\bm{\Lambda}) &= \frac{det\left\lbrace(\textbf{X}^{\prime}\textbf{X})^{-1}\right\rbrace^{1/2}det\left\lbrace \left((\textbf{I}-\textbf{P})\bm{\Sigma}_{\nu}(\textbf{I}-\textbf{P})+\textbf{I}\right)^{-1}\right\rbrace^{1/2} f(\bm{\Lambda})}{\kappa^{(n-p)/2 + \alpha}(2\pi)^{(n-p)/2} (\frac{(2\alpha-p)\textbf{y}^{\prime}\left((\textbf{I}-\textbf{P})\bm{\Sigma}_{\nu}(\textbf{I}-\textbf{P})+\textbf{I}\right)^{-1}\textbf{y}/(2\kappa)}{2\alpha-p}+ 1)^{(n-p)/2+\alpha}}.
\end{align}
\noindent
This implies that
\begin{equation*}
	f(\textbf{y}\vert\textbf{X},\bm{\Lambda})\propto \mathcal{MT}(\bm{0}_{n,1},\bm{\Sigma}_{n}, 2\alpha - p),
\end{equation*}
where $\bm{\Sigma}_{n} = \frac{2\kappa}{(2\alpha-p)} \left((\textbf{I}-\textbf{P})\bm{\Sigma}_{\nu}(\textbf{I}-\textbf{P})+\textbf{I}\right)$. It follows from standard properties of the multivariate t-distribution that \citep{ding2016conditional},
\begin{align}
	\nonumber
	\textbf{y}_{m}\vert \textbf{y}_{o},\bm{\Lambda},\bm{\Sigma}_{\nu}&\sim \mathcal{MT}(\bm{\Sigma}_{m,o}\bm{\Sigma}_{o}^{-1}\textbf{y}_{o},\rho(\bm{\Sigma}_{m}-\bm{\Sigma}_{m,o}\bm{\Sigma}_{o}^{-1}\bm{\Sigma}_{o,m}), 2\alpha + n_{o}-p)\\
	\label{eq:multivt}
	\textbf{y}_{o}\vert \bm{\Lambda},\bm{\Sigma}_{\nu} &\sim \mathcal{MT}(\bm{0}_{n_{o}},\bm{\Sigma}_{o}, 2\alpha - p).
\end{align}
This completes the result.\\

\noindent
\textit{Proof of the expressions of $f(vec(\bm{\Lambda})\vert \textbf{y}_{o})$ in Equation (\ref{eq:post}):} From (\ref{eq:multivt}) and the fact that $\bm{\Sigma}_{o} = \frac{2\kappa}{(2\alpha-p)} \textbf{B}\bm{\Lambda}\textbf{B}^{\prime}$ we have,
\begin{align*}
	f(\textbf{y}_{o},\textbf{X},\bm{\Lambda}) &\propto \frac{\left(\prod_{i = 1}^{n_{o}}(\lambda_{i}+1)\right)^{-1/2} f(\bm{\Lambda})}{ (\frac{1}{2\alpha - p}\frac{2\alpha-p}{2\kappa}\textbf{y}_{o}^{\prime}\textbf{B}\bm{\Lambda}\textbf{B}^{\prime}\textbf{y}_{o}/2+ 1)^{(n_{o}-p)/2+\alpha}}\\
	&= \frac{\left(\prod_{i = 1}^{n_{o}}(\lambda_{i}+1)\right)^{-3/2} f(\bm{\Sigma}_{\nu}\vert \bm{\Lambda})}{ (\frac{1}{2\alpha-p}\sum_{i = }^{n_{o}}\frac{h_{i}^{2}}{\lambda_{i}+1}+ 1)^{(n_{o}-p)/2+\alpha}}\\
	\propto
	&f(vec(\bm{\Lambda})\vert \textbf{y}_{o}),
\end{align*}
\noindent
where recall $\textbf{h} = (h_{1},\ldots,h_{n_{o}})^{\prime} = \left(\frac{2\alpha - p}{2\kappa}\right)^{1/2}\textbf{B}^{\prime}\textbf{y}_{o}$.\\

\noindent
\textit{Proof of Proposition 4:} Consider $\textbf{g} = (g_{1},\ldots, g_{n_{o}})^{\prime}$ with density
\begin{equation}
	f_{g}(\textbf{g}\vert \textbf{D}) = \mathcal{MT}(\bm{0}_{n_{o}}, \textbf{D},\gamma)\frac{I(-1<g_{1}<1,\ldots, -1<g_{n}<1)}{P_{g}}
\end{equation}
\noindent
 where $\textbf{D} = diag(d_{1}^{2},\ldots, d_{n_{o}}^{2})$ and
 \begin{equation*}
P_{g}=\int_{-1}^{1}\ldots\int_{-1}^{1}\mathcal{MT}(\bm{0}_{n_{o}}, \textbf{D},\gamma)dg_{1}\ldots dg_{n_{o}}.
 \end{equation*} 
 \noindent
Partition the parameter space $\mathcal{A} = \{(g_{1},\ldots, g_{n}): f(\textbf{g}\vert \textbf{D})>0\}$ into the sets $\mathcal{A}_{1},\ldots, \mathcal{A}_{2^{n_{o}}}$, where $\mathcal{A}_{1} = \{(g_{1},\ldots, g_{n}): g_{1} \in (-\infty,0), g_{2}\in \left[\right. 0, \infty),\ldots,  g_{n_{o}}\in \left[\right. 0, \infty)\}$ and the remaining subsets $\mathcal{A}_{i}$ consider every combination of restricting components of $\textbf{g}$ to be either negative or non-negative. Consider the transformation $q_{i} = \frac{1}{g_{i}^{2}}-1$, which produces the mapping $w_{ij}(g_{i}) = \frac{1}{g_{i}^{2}}-1$, inverse mapping $w_{ij}^{-1}(g_{i}) = \left(\frac{1}{q_{i}+1}\right)^{1/2}$, and Jacobian $J_{j} = \prod_{i = 1}^{n_{o}}\frac{-r_{ij}}{2} \left(\frac{1}{q_{i}+1}\right)^{3/2}$, where $r_{ij} = 1$ if $g_{i}\ge 1$ for $g_{i}\in A_{j}$ and $r_{ij} = -1$ if $g_{i}\le -1$ for $g_{i}\in A_{j}$. Then it follows from standard change of variables that \citep[][pg. 185]{casellaBerger},
\begin{align*}
	f({q}_{1},\ldots, {q}_{n_{o}}\vert \textbf{D}) &= \sum_{j = 1}^{2^{n_{o}}} f_{g}(w_{1j}^{-1}(g_{1}),\ldots, w_{n_{o}j}^{-1}(g_{n_{o}})\vert \textbf{D})|J_{j}|\\
	&=\sum_{j = 1}^{2^{n_{o}}} \frac{\Gamma(\frac{\gamma+n_{o}}{2})}{\Gamma(\gamma/2)\gamma^{n_{o}/2}(\prod_{i = 1}^{n_{o}}d_{i}^{2})^{1/2}} \frac{1}{\left(\frac{1}{\gamma}\sum_{k = 1}^{n_{o}}\frac{1/d_{k}^{2}}{q_{k}+1}+1\right)^{\frac{n_{o}+\gamma}{2}}} \prod_{i = 1}^{n_{o}}\frac{1}{2}\left(\frac{1}{q_{i}+1}\right)^{3/2}\\
	&=2^{n_{o}}\frac{\Gamma(\frac{\gamma+n_{o}}{2})}{\Gamma(\gamma/2)\gamma^{n_{o}/2}(\prod_{i = 1}^{n_{o}}d_{i}^{2})^{1/2}} \frac{1}{\left(\frac{1}{\gamma}\sum_{k = 1}^{n_{o}}\frac{1/d_{k}^{2}}{q_{k}+1}+1\right)^{\frac{n_{o}+\gamma}{2}}} \prod_{i = 1}^{n_{o}}\frac{1}{2}\left(\frac{1}{q_{i}+1}\right)^{3/2}\\
	&=\frac{\Gamma(\frac{\gamma+n_{o}}{2})}{\Gamma(\gamma/2)\gamma^{n_{o}/2}(\prod_{i = 1}^{n_{o}}d_{i}^{2})^{1/2}} \frac{1}{\left(\frac{1}{\gamma}\sum_{k = 1}^{n_{o}}\frac{1/d_{k}^{2}}{q_{k}+1}+1\right)^{\frac{n_{o}+\gamma}{2}}} \prod_{i = 1}^{n_{o}}\left(\frac{1}{q_{i}+1}\right)^{3/2},
\end{align*}
\noindent
where $|-r_{ij}| = 1$. Thus, to simulate from $f(vec(\bm{\Lambda})\vert \textbf{y}_{o})$ first simulate $\textbf{g}$ from a $f_{g}$. Then $\lambda_{i}$ is equal in distribution to $1/g_{i}^{2} - 1$ for $i = 1,\ldots, n_{o}$.\\

\section*{Appendix B: Orthogonality Is Not Equivalent to the RSR}\label{sec:counter-example}

One might be tempted to claim that our adjusted model is simply reversing the process of orthogonalization, as the definition of $\bm{\beta}^{adj}$ simply reverses the process of orthogonality. However, this is not the case because orthogonalization holds under any prior distributional assumptions, and our result on the posterior predictive adjustment expressly needs the assumption of location-invariant prior distributions. That is, Conclusions 2 and 3 can fail if different prior distributions are assumed. We provide an example of such a case in this section. 

Refer to the SLMM with a normal prior on $\bm{\beta}$ as,
\begin{align}
	\nonumber 
	f_{SLMM}(\textbf{y}\vert \textbf{X}, {\bm{\beta}},\bm{\nu},\sigma^{2}) & = N(\textbf{X}{\bm{\beta}}+\bm{\nu},\sigma^{2}\textbf{I})\\
	\nonumber
	f_{n}({\bm{\beta}}\vert \sigma_{\beta}^{2},\bm{\nu},\textbf{X})  &= N(\bm{0}_{p},\sigma_{\beta}^{2}\textbf{I}_{p})\\
	\nonumber
	f_{SLMM}(\bm{\nu}\vert \bm{\Sigma}_{\nu},\sigma^{2}) &= N(\bm{0}_{n},\bm{\Sigma}_{\nu})\\
	\nonumber
	&f(\bm{\Sigma}_{\nu})\\
	\label{eq:nslmm:nointer}
	&f(\sigma^{2}),
\end{align}
\noindent
where the subscript ``n'' in $f_{n}$'' stands for a ``normal.'' This case is important as it is common in the RSR literature to assume this prior \citep{hughes,bradleyMSTM}.

Consider the misspecified augmented RSR (i.e., it is incorrectly assumed $\bm{\beta}_{RSR} = \bm{\delta}_{RSR}$) with a normal prior assumption for $\bm{\beta}$ (nRSR):
\begin{align}
	\nonumber
	f_{aRSR}(\textbf{y}\vert  \bm{\beta}_{RSR},\bm{\nu},\sigma^{2})&= N(\textbf{X}\bm{\beta}_{RSR} + (\textbf{I}-\textbf{P})\bm{\nu},\sigma^{2}\textbf{I})\\
	\nonumber
	f_{n}(\bm{\beta}_{RSR}\vert \sigma_{\beta}^{2})  &= N(\bm{0}_{p},\sigma_{\beta}^{2}\textbf{I}_{p})\\
	\nonumber
	f_{SLMM}(\bm{\nu}\vert \bm{\Sigma}_{\nu}) &= N(\bm{0}_{n},\bm{\Sigma}_{\nu})\\
	\nonumber
	&f(\bm{\Sigma}_{\nu})\\
	\label{eq:rsr:normalbhm}
	&f(\sigma^{2}),
\end{align}
\noindent
which produces the joint distribution,
\begin{align*}
	& f_{nRSR}(\textbf{y}, \bm{\beta}_{RSR},\bm{\nu},\bm{\Sigma}_{\nu},\sigma^{2})\\
	& = f_{aRSR}(\textbf{y}\vert \textbf{X}, \bm{\beta}_{RSR},\bm{\nu},\bm{\Sigma}_{\nu},\sigma^{2})	f_{n}(\bm{\beta}_{RSR}\vert \textbf{X},\bm{\nu},\bm{\Sigma}_{\nu},\sigma^{2}),\sigma^{2})f_{SLMM}(\bm{\nu}\vert \textbf{X},\bm{\Sigma}_{\nu},\sigma^{2})f(\bm{\Sigma}_{\nu}\vert \textbf{X},\sigma^{2})f(\sigma^{2}\vert \textbf{X}).
\end{align*}
\noindent
Now apply our posterior predictive adjustment to the misspecified aRSR via the change-of-variables in (\ref{eq:c-o-v}) to produce the adjusted model,
\begin{align}
	\label{eq:fail:rsr:toslmm}
	&f_{nRSR}(\textbf{y},\bm{\beta}^{adj},\widetilde{\bm{\nu}},\bm{\Sigma}_{\nu},\sigma^{2}) = f_{SLMM}(\textbf{y}\vert  \bm{\beta}^{adj},\widetilde{\bm{\nu}},\sigma^{2})f_{n}(\bm{\beta}^{adj}\vert \widetilde{\bm{\nu}},\sigma_{\beta}^{2})f_{SLMM}(\widetilde{\bm{\nu}}\vert \bm{\Sigma}_{\nu},\sigma^{2})f(\bm{\Sigma}_{\nu})f(\sigma^{2}),
\end{align}
where $f_{n}(\bm{\beta}^{adj}\vert \widetilde{\bm{\nu}},\sigma_{\beta}^{2}) = N(-(\textbf{X}^{\prime}\textbf{X})^{-1}\textbf{X}^{\prime}\widetilde{\bm{\nu}},\sigma_{\beta}^{2}\textbf{I}_{p})$. Equation (\ref{eq:fail:rsr:toslmm}) can be expressed hierarchically as
\begin{align}
	\nonumber 
	f_{SLMM}(\textbf{y}\vert \textbf{X}, \bm{\beta}^{adj},\widetilde{\bm{\nu}},\sigma^{2}) & = N(\textbf{X}\bm{\beta}^{adj}+\widetilde{\bm{\nu}},\sigma^{2}\textbf{I})\\
	\nonumber
	f_{n}(\bm{\beta}^{adj}\vert \sigma_{\beta}^{2},\widetilde{\bm{\nu}},\textbf{X})  &= N(-(\textbf{X}^{\prime}\textbf{X})^{-1}\textbf{X}^{\prime}\widetilde{\bm{\nu}},\sigma_{\beta}^{2}\textbf{I}_{p})\\
	\nonumber
	f_{SLMM}(\widetilde{\bm{\nu}}\vert \bm{\Sigma}_{\nu},\sigma^{2}) &= N(\bm{0}_{n},\bm{\Sigma}_{\nu})\\
	\nonumber
	&f(\bm{\Sigma}_{\nu})\\
	\label{eq:nslmm}
	&f(\sigma^{2}).
\end{align}
\noindent
We again see that in the data model, $\textbf{X}\bm{\beta}^{adj}$ is interpreted as a linearly dependent regression effect in the likelihood. However, the adjusted misspecified aRSR model in (\ref{eq:nslmm}) is different from that of the nSLMM in (\ref{eq:nslmm:nointer}). Consequently, Conclusion 3 does not hold for a normal prior on ${\bm{\beta}_{RSR}}$, and the corresponding equivalence relationships in Proposition 1 do not hold. 

In a similar manner, if we express the orthogonal reparameterization of the SLMM stated in Equation (2) hierarchically (with a normal distribution on $\bm{\beta}$), we obtain,
\begin{align}
	\nonumber
	f_{aRSR}(\textbf{y}\vert  \bm{\delta},\widetilde{\bm{\nu}},\sigma^{2})&= N(\textbf{X}\bm{\delta} + (\textbf{I}-\textbf{P})\widetilde{\bm{\nu}},\sigma^{2}\textbf{I})\\
	\nonumber
	f_{n}(\bm{\delta}\vert \widetilde{\bm{\nu}},\sigma_{\beta}^{2})  &= N((\textbf{X}^{\prime}\textbf{X})^{-1}\textbf{X}^{\prime}\widetilde{\bm{\nu}},\sigma_{\beta}^{2}\textbf{I}_{p})\\
	\nonumber
	f_{SLMM}(\widetilde{\bm{\nu}}\vert \bm{\Sigma}_{\nu}) &= N(\bm{0}_{n},\bm{\Sigma}_{\nu})\\
	\nonumber
	&f(\bm{\Sigma}_{\nu})\\
	\label{eq:slmm:normalbhm}
	&f(\sigma^{2}),
\end{align}
\noindent
which is not the same as the aRSR model in (\ref{eq:rsr:normalbhm}) that would replace $(\textbf{X}^{\prime}\textbf{X})^{-1}\textbf{X}^{\prime}\widetilde{\bm{\nu}}$ with $\bm{0}_{p}$. Consequently, Conclusion 2 does not hold in this case, and the corresponding equivalence relationships in Proposition 1 again do not hold.

\section*{Appendix C: Additional Simulation Results} We provide additional simulation results with $n = 200$, signal to noise ratio equal to 10, and $\rho =0.5$. See Section 5.1 for a discussion of these results.

\begin{figure}[htbp]
	\centering
	\includegraphics[width=0.8\textwidth]{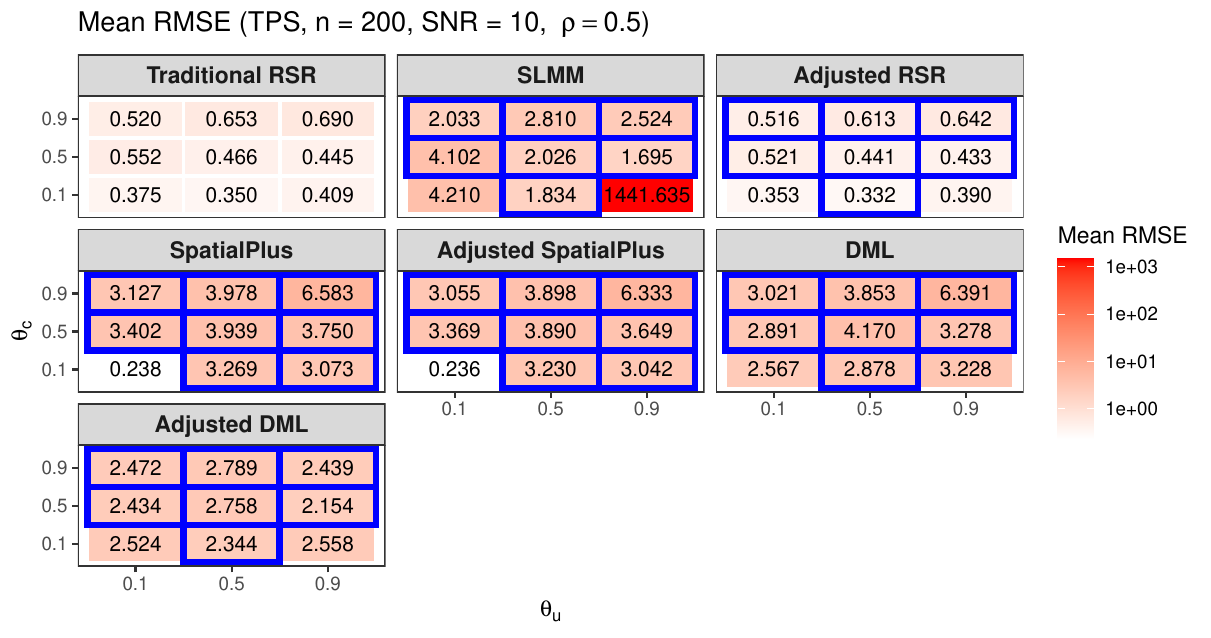}
	\caption{The average $\mathrm{RMSE}_{\beta}$ over 50 simulated replicate datasets at $n = 200$, SNR = 10, $\rho = 0.5$, and SLMM, RSR, and adjusted RSR implemented with thin-plate splines. The $y$-axis considers different values of $\theta_{c}$, and the $x$-axis considers different values of $\theta_{u}$. All other simulation specifications are listed in Section~\ref{sec:verify}. The panel title ``Adjusted RSR'' corresponds to estimates derived from the posterior mean of $\bm{\beta}_{trn}^{adj}$ setting $\bm{\mu}_{\beta}$ equal to the OLS estimator. The panel title ``Adjusted SpatialPlus'' corresponds to estimates derived from the posterior mean of $\bm{\beta}_{trn}^{adj}$ setting $\bm{\mu}_{\beta}$ equal to the Spatial+ estimator. The panel title ``Adjusted DML'' corresponds to estimates derived from the posterior mean of $\bm{\beta}_{trn}^{adj}$ setting $\bm{\mu}_{\beta}$ equal to the DML shift estimator.  Blue cells indicate that a pairwise $t$ test indicates a significant difference between the respective SLMM/Adjusted RSR, SpatialPlus/Adjusted SpatialPlus and DML/Adjusted DML panels.}
	\label{fig:rmse1}
\end{figure}

\begin{figure}[htbp]
	\centering
	\includegraphics[width=0.8\textwidth]{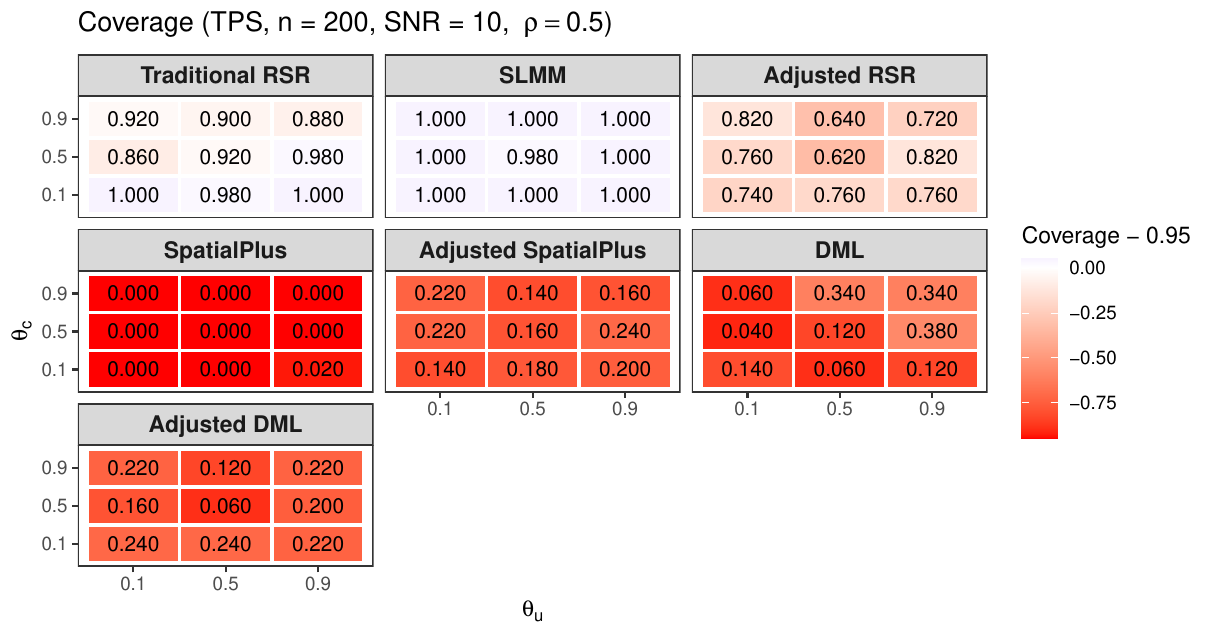}
	\caption{The coverage of $\bm{\beta}$ over 50 simulated replicate datasets at $n = 200$, SNR = 10, $\rho = 0.5$, and SLMM, RSR, and adjusted RSR implemented with thin-plate splines. The $y$-axis considers different values of $\theta_{c}$, and the $x$-axis considers different values of $\theta_{u}$. All other simulation specifications are listed in Section~\ref{sec:verify}. The panel title ``Adjusted RSR'' corresponds to estimates derived from the posterior mean of $\bm{\beta}_{trn}^{adj}$ setting $\bm{\mu}_{\beta}$ equal to the OLS estimator. The panel title ``Adjusted SpatialPlus'' corresponds to estimates derived from the posterior mean of $\bm{\beta}_{trn}^{adj}$ setting $\bm{\mu}_{\beta}$ equal to the Spatial+ estimator. The panel title ``Adjusted DML'' corresponds to estimates derived from the posterior mean of $\bm{\beta}_{trn}^{adj}$ setting $\bm{\mu}_{\beta}$ equal to the DML shift estimator. }
	\label{fig:coverage1}
\end{figure}

\begin{figure}[htbp]
	\centering
	\includegraphics[width=0.8\textwidth]{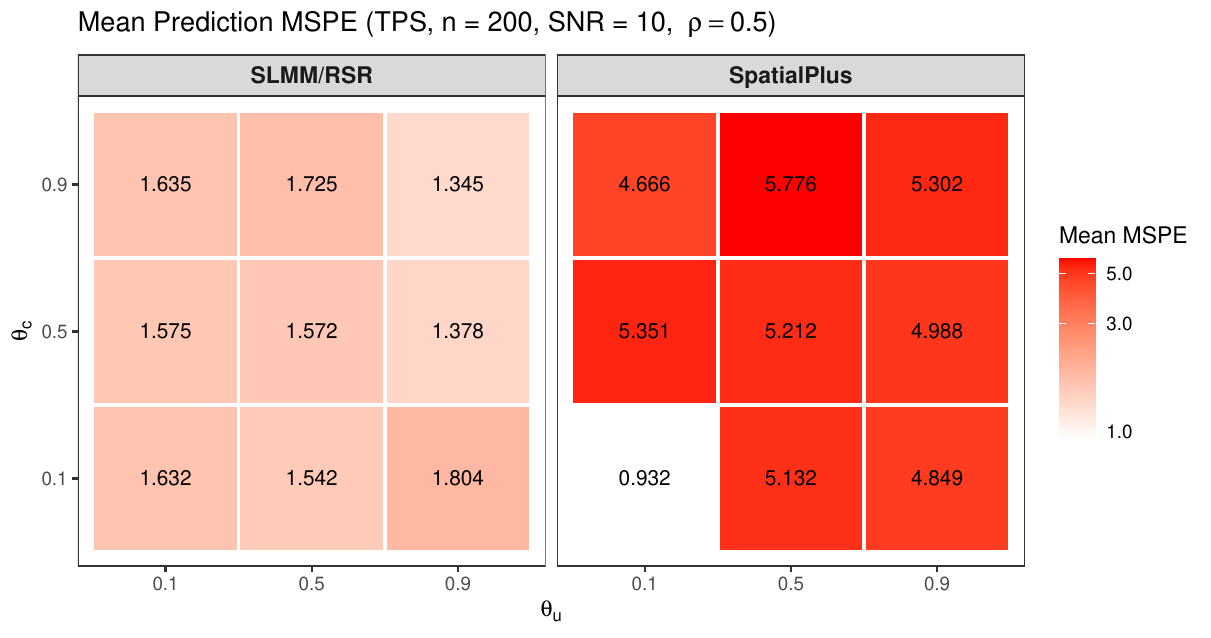}
	\caption{The average $\mathrm{RMSPE}$ over 50 simulated replicate datasets at $n = 200$, SNR = 10, $\rho = 0.5$, and SLMM, RSR, and adjusted RSR implemented with thin-plate splines. The $y$-axis considers different values of $\theta_{c}$, and the $x$-axis considers different values of $\theta_{u}$. All other simulation specifications are listed in Section~\ref{sec:verify}. The panel title ``Adjusted RSR'' corresponds to estimates derived from the posterior mean of $\bm{\beta}_{trn}^{adj}$ setting $\bm{\mu}_{\beta}$ equal to the OLS estimator. The panel title ``Adjusted SpatialPlus'' corresponds to estimates derived from the posterior mean of $\bm{\beta}_{trn}^{adj}$ setting $\bm{\mu}_{\beta}$ equal to the Spatial+ estimator. The panel title ``Adjusted DML'' corresponds to estimates derived from the posterior mean of $\bm{\beta}_{trn}^{adj}$ setting $\bm{\mu}_{\beta}$ equal to the DML shift estimator.  All respective cells produce a pairwise $t$-test that is significant.}
	\label{fig:rmspe2}
\end{figure}

\end{document}